\documentclass[11pt,times,onecolumn,peerreview]{IEEEtran}
\usepackage{eucal}
\usepackage{times}
\usepackage{latexsym}
\usepackage{amsmath,epsfig,amssymb}

\IEEEoverridecommandlockouts

\begin{document}
\title{Mobile Wireless Localization through Cooperation}

\author{Xingkai Bao, ~\IEEEmembership{Member,~IEEE,}  and Jing Li (Tiffany)
~\IEEEmembership{Senior member,~IEEE,}\thanks{This research is supported by the National Science Foundation under the Grant No. CCF-0635199, CCF-0829888, CMMI-0928092, and OCI-1133027. Bao and Li are
with the department of electrical and computer engineering, Lehigh
University, Bethlehem, PA 18015. Email: \{xib3,
jingli\}@ece.lehigh.edu.}
}
\maketitle

\begin{abstract}
This paper considers $N$ mobile nodes that move together in the vicinity of each other, whose
initial poses as well as subsequent movements must be accurately
tracked in real time with the assist of $M(\geq 3)$ reference nodes. By
engaging the neighboring mobile nodes in a simple but effective
cooperation, and by exploiting both the time-of-arrival
(TOA) information (between mobile nodes and reference nodes) and
the received-signal-strength (RSS) information (between mobile
nodes), an effective new localization strategy, termed ``cooperative
TOA and RSS'' (COTAR), is developed. An optimal maximum likelihood detector is first formulated, followed by the derivation of a low-complexity iterative approach that can practically achieve the Cramer-Rao lower bound.
Instead of using simplified channel models as in many previous studies, a sophisticated and realistic channel model is used, which can effectively account for the critical fact that the direct path is not necessarily the strongest path. Extensive simulations are conducted in static and mobile settings, and various practical issues and system parameters are evaluated. It is shown that COTAR significantly outperforms the existing strategies, achieving a localization accuracy of only a few tenths of a meter in clear environments and a couple of meters in
heavily obstructed environments.
\end{abstract}

\begin{keywords}
mobile localization, wireless localization, user cooperation, maximum likelihood estimation, iterative detection, time-of-arrival, received-signal-strength
\end{keywords}

\section{Introduction}
Efficient and accurate localization is desirable for a variety of
wireless ad-hoc networks, especially {\it mobile} ad hoc networks
and robot systems that will be forming dynamic coalitions for
important missions like search-and-rescue and disaster recovery
\cite{new2}-\cite{Predict_TOF}. While the world-wise deployment of
global positioning systems (GPS) has significantly alleviated the
localization problem \cite{GPS}, there exist several scenarios, such
as inside a building or in a forest,  where GPS signal is lacking.
Further, because of the hardware cost, size, battery and other
concerns, not all the nodes in a system may be equipped with a GPS
receiver.

Wireless localization is being studied in the system architecture, protocol,
and algorithm level. This paper considers mobile localization in the
algorithm level, and attacks {\it position tracking}, the
fundamental computation problem at the heart of mobile localization.
In some situations, the initial pose of the mobile node  is known
and hence only incremental errors need to be compensated. The more
challenging case, also referred to as the ``global localization
problem,'' is when  the initial position and all of the subsequent
movements must be detected from the scratch \cite{mobile_loc1}. A
desirable mobile localization algorithm should provide good accuracy
and robustness. The former is particularly relevant to indoor
applications where localization errors should in general be
controlled below a few of meters or even a few tenths of a meter.
The latter is pivotal to the system survivability in a dynamic environment
where the communication channels may be randomly faded, shadowed, or heavily obstructed
without line-of-sight (LOS). Finally, the good quality of localization should be achieved
in a cost-effective manner with simple hardware, since many mobile nodes,
especially those in large systems, may be constrained by their cost and sizes,
and hence are only provisioned with limited resources like a single
small omnidirectional antenna, an inexpensive
low-precision clock and a small piece of bandwidth. This paper aims to develop new localization strategies
that can achieve all of these goals.

A fundamental method to determine the point location of a target
node (i.e. a robot or a cell phone) is triangulation. Existing
techniques exploit different modalities of the radio frequency (RF)
signal, such as received angle-of-arrival (AOA), received signal
strength (RSS), time-of-flight (TOF) (also known as time-of-arrival
or TOA), and time-difference-of-arrival (TDOA), to compute and
estimate the location or the relative location of the {\it target
node} with respect to the {\it
reference nodes} or {\it anchor nodes}\cite{new2,new1}.

Since the AOA technique requires the availability of expensive
directional antennas at the reference nodes and/or at the sensor
nodes, this is much less employed in practical systems than the
other techniques. The RSS technique exploits the received signal
power together with a path-loss and shadowing model to provide a
distance estimate between the sender and the receiver. It works
effectively when the nodes are close to each other, but is quite
sensitive to shadowing, multi-path fading and scattering, and hence
the performance deteriorates quickly with the increase of the
distance \cite{RSS_limit}. The TOA technique uses the signal
propagation time between a target node to a reference node to
compute the target position  in a circular track. The TDOA
technique estimates the distance difference of a target node to two
reference nodes through received-time difference, thus determining
the position of the target in an ellipse track without knowledge of
the signal transmitting time. Comparing to the RSS technique, the
TOA/TDOA-based methods can achieve a reasonable accuracy at a
much larger measurement range. It is also possible to improve the
accuracy of TOA/TDOA by increasing the signal transmit power and the
system bandwidth, but since the localization inaccuracy is largely caused
by multi-path and no-line-of-sight (NLOS), the TOA/TDOA
accuracy will hit a bottleneck at some point, beyond which further
increasing transmit power or bandwidth does not help
\cite{Predict_TOF}. To combat NLOS and other channel dynamics and
to improve positioning accuracy, hybrid schemes that employ multiple different modalities
are proposed \cite{TDOA-GRA}\cite{Hybrid_TDOA_AOA}. For estimating the
incremental pose change of a mobile node, {\it a posteriori}
methods, including Kalman filters, multi-hypothesis Kalman filters,
Markov localization and Monte Carlo location
\cite{robot2}\cite{mobile_loc1}, are also shown to be useful.

This paper considers the global localization problem in a highly dynamic environment
where the communication channels may be randomly faded, shadowed, or heavily obstructed without line-of-sight (LOS).
One big challenge about timing an RF signal in an obstructed environment is that the direct path signal is not
necessarily the strongest path, and hence the detector tends to miss the direct path signal but catch the strongest path signal,
causing the measurement of the propagation time to be longer than it actually is\footnote{It should be noted that this type of error is different from the usual measurement error which can be modeled by a Gaussian random variable and which can in theory cancel out, given enough repeated measures.}.
However, the majority of the existing studies have only looked into clear-sight environments,
and hence have not explicitly accounted for this critical source of error. In this paper, we
will adopt a sophisticated and realistic 2-ray channel model which is derived from
real-world experiments and which can effectively capture the reflective path distortion \cite{Predict_TOF},
in addition to all the other  usual noise and measurement error.

Unlike the majority of the work that considers localization as independent tasks of individual (mobile) nodes,
here we consider {\it cooperative localization}.
Many applications involve locating and tracking multiple target
nodes that stay in the vicinity of each other. Examples include
tracking first-responders operating in a buddy system in a search-and-rescue
task in a building, positioning a group of scientists exploring a
forest, or monitoring a robot that is equipped with two (or more)
sensors on both sides of the body performing tasks in a cave.
Exploring practical application needs, we propose to engage
neighboring nodes in a simple but effective cooperation, and to jointly
exploit the TOA information (between each target node and each reference node)
and the RSS information (between neighboring target nodes) to accurately localize all the target nodes at once.
The proposed strategy, thereafter referred to as {\it cooperative TOA and RSS} (COTAR), is
advantageous in several ways:

\begin{itemize}
\item COTAR operates on simple hardware and limited resources
such as simple omni-directional antennas (rather than expensive directional
antennas), and a narrow bandwidth (e.g. 2M Hz as in Zigbee sensors) rather than ultra-wide band.

\item Rather than treat sensors as independent nodes,
COTAR makes clever use of the vicinity of two, three or more
target nodes, and introduces among them very simple, practical and
effective cooperation. The cooperation increases the signal spatial diversity and the geometric diversity, resulting in a significant improvement in the localization accuracy at the cost of only marginal overhead.

\item There has been consideration of jointly utilizing TOA and RSS in the literature \cite{TDOA-GRA}\cite{Hybrid_TDOA_AOA},
but the strategies thereof employ TOA and RSS in a rather natural and straight-forward manner, namely, each target node individually sends a beacon signal such that the $M$ reference nodes record both RSS and TOA information and use all of the $2M$ measures to localize this target. By combining RSS with TOA, such hybrid schemes help improve the localization accuracy in short ranges, but as soon as the measurement range increases beyond 10 meters, they perform no better than TOA-only systems (yet with a higher complexity) \cite{TDOA-GRA}\cite{Hybrid_TDOA_AOA}.
In comparison, COTAR combines TOA and RSS in a more clever manner and exploits each in its most favorable way -- TOA for long ranges and RSS for short ranges, and hence achieves a significant performance gain in both short and long ranges. Additionally, COTAR can also effectively combat various channel conditions (e.g. fading, shadowing and NLOS) and measurement distortions, and is particularly suitable for heavily
obstructed environments such as in an office building or in a
forest.

\item The proposed strategy is rather general, allowing the joint localization
of $N$ (where $N$ can be any number $\ge 2$) cooperative target
nodes by exploiting $M$ (where $M$ can be any number $\ge 3$)
reference nodes and two radio signal measurements (TOA and RSS).
While the exact formulation takes a rather sophisticated, albeit
systematic, nonlinear form, we have introduced a way to simplify it
by approximating it to a low-complexity linear problem that can be
iteratively refined in an efficient manner. Linearity makes
incremental position tracking effective, and iterative refinement
makes estimation accurate.

\item The proposed strategy is also analytically tractable.
Asymptotic lower bounds on the localization distortion, quantified by
the root mean square (RMS) distortion and the Cramer-Rao bound (CRB), are derived for COTAR, and simulations
yield results extremely close to the theoretical bounds.
\end{itemize}

Extensive simulations are conducted in static and mobile settings and in clear-sight and obstructed environments, and various practical issues and system parameters are evaluated, including the impact of the missing RSS information, the number of collaborating target nodes, the number and geometry of the reference nodes, the measurement range, and the mobility speed. Comparison to the conventional RSS, TOA, and TOA/RSS hybrid schemes confirm the superiority of the proposed COTAR scheme.

The remainder of the paper is organized as follows. Section \ref{sec:systemModel} introduces the system model. Section \ref{sec:protocol} discusses the proposed COTAR localization protocol. Section \ref{sec:algorithm} formulates the maximum likelihood detection method and discusses its low-complexity linear approximation with iterative refinement. Section \ref{sec:analysis} analyzes the performance of the proposed strategy and presents a lower bound on root mean square error. Section \ref{sec:simulation} presents and discusses the simulation results. Finally Section \ref{sec:conclusion} concludes the paper.

\section{System Model}
\label{sec:systemModel}

Without loss of generality, we consider employing $M$ ($M\geq 3$)
reference nodes with pre-known coordinates vector
$(\mathbf{x}_r,\mathbf{y}_r)$ to help localize a cluster of $N$
target nodes with unknown coordinate vector
$(\mathbf{x},\mathbf{y})$ in a two-dimensional space.
In the sequel, unless otherwise stated, we will use bold fonts (e.g.
$\mathbf{x}$) to denote vectors or matrices, and use regular fonts
(e.g. $x$) to denote scalars. All the vectors are by default column
vectors. Further, subscript $s$ and $r$ denote the quantities
associated with the targets nodes and the reference nodes,
respectively.

\subsection{Path-Loss Model and RSS Estimate Accuracy}

The average path loss of the radio signal propagation in general
follows an exponential attenuation model, and can be expressed in unit
of dB as:
\begin{align}\label{eqn:attenuation}
\bar{g}(d)=10\eta\log_{10}(d)+g_0,
\end{align}
where $\eta$ is the attenuation factor, $d$ is the distance between
the transmitter and the receiver, and $g_0$ is the calibration
pass-loss. The shadowing, multi-path and scattering phenomenons are
collectively modeled as Gaussian distributed noise in dB (sometimes
referred to as logarithmic Gaussian noise in the literature) with
variance $\sigma_g^2$. Thus the individual path-loss or signal
attenuation follows a Gaussian distribution (when measured in dB):
\begin{align}\label{eqn:Gaussian_RSS}
g(d)=\mathcal{N}(\bar{g}(d),\sigma^2_g).
\end{align}
Unless otherwise specified, we will use the attenuation factor
$\eta=3.086$ and the standard deviation of the path-loss
$\sigma_g=8$ dB in the analysis and the simulations.

From (\ref{eqn:attenuation}) and (\ref{eqn:Gaussian_RSS}), we can
calculate the RSS estimation accuracy, measured by the standard
derivation of the distance estimate based on RSS, as follows:
\begin{align}
\mathbf{std}(\hat{d})=\frac{\ln(10) \sigma_g d}{10\eta}.
\end{align}
This clearly indicates that the localiztion accuracy of RSS-based techniques increases linearly with the measurement range $d$.

\subsection{Multi-Path Model and TOA Estimate Accuracy}

To study the TOA measurement error, consider modeling the
multi-path channel profile as:
\begin{align}
h(\tau)=\sum_{k=0}^{L_p-1}\alpha_k\delta(\tau-\tau_k),
\end{align}
where $L_p$ is the number of multi-path, and $\tau_k$ and $\alpha_k$
are the time delay and the attenuation coefficient of the $k$th
multi-path, respectively. To be practical, we assume that the
receivers use a simple early-late structure and determine the signal
arrival time by the first zero-crossing point of the correlator's
output \cite{Predict_TOF}. One major challenge about measuring
TOA, especially in a highly obstructed environment with
severe reflection and multi-path, is that the direct path (i.e. the
first arrival path) is not necessarily the strongest path.
As such, the receiver will likely miss the direct path, catch the strongest path and mistaken it for
the direct path, thus arriving at a longer TOA estimate (and a
longer distance estimate) than it actually is\footnote{Note that 1
ns of TOA error amounts to 0.3 meter of distance error.}. To account
for this factor in our study, we use the two-ray approximation
model developed in \cite{Predict_TOF} to capture and simulate the
effect of the strongest path. The two-ray approximation model consider two dominant paths: the direct path that occurs at some refrence time
$0$, and a second path that is usually a ground-reflected path (may be the strongest path) occurring after a small delay at time $\bar{\tau}$. The small time dealy $\bar{\tau}$ is known as the \emph{mean excess delay} \cite{Predict_TOF}. The channel coefficient $h$ is thus mathematically
expressed as follows:
\begin{align}
h(\tau)=\sqrt{\frac{K}{K+1}}\,\delta(0)+\sqrt{\frac{1}{K+1}}\,\alpha\delta(\bar{\tau}),
\end{align}
where $\alpha$ is the Rayleigh faded random coefficient, and $K$ is termed the Rician factor in the literature, which denotes the direct-path to multi-path energy ratio. A large $K$ indicates that the channel is clear sight having strong LOS, while a small $K$ indicates heavy obstruction and severe multi-path.

Following Kim {\it et. at.}'s measurements in realistic engineering buildings \cite{Pulse_propagation}, we consider two representative cases for the localization environment: the case of clear line-of-sight, which is associated with a mean excess delay $\bar{\tau}=25.8$ ns and a Rician factor $K=5$, and the case of heavy obstruction, which has $\tau=76.9$ns and  $K=2$. Additionally, following the performance predicting techniques in \cite{Predict_TOF}, we arrive at the the TOA estimation errors for different situations, which are listed in Table.\ref{tab:TOA_std} and which will be used in our analysis and simulations.
\begin{table}[h]
\vspace{-0.2cm}
\centering
\caption{TOA estimation error standard derivation for different localization environments}
\vspace{-0.5cm}
\begin{tabular}{|c|c|c|c|}
  \hline
  situation                     & Rician factor & mean excess delay (ns) & time-of-arrival error std (ns)\\\hline
  clear line-of-sight  & 5 & 25.8 & 8.8 \\\hline
  heavily obstructed   & 2 & 76.9 & 40.2 \\
  \hline
\end{tabular}\label{tab:TOA_std}
\vspace{-0.8cm}
\end{table}

\section{Cooperative TOA and RSS Localization Protocol}\label{sec:protocol}
The proposed COTAR strategy engages the set of target nodes in a
simple but effective cooperation to facilitate each other's
positioning. Consider the target nodes sending beacon packets one by
one in a sequential manner. Due to the wireless broadcast advantage,
when one target node transmits beacon signals, all the other target
nodes in the vicinity can hear and measure the
received-signal-strength from the transmitting node rather
accurately. These RSS measurements contain valuable information
about the relative distances and geological topology of the set of
target nodes, which, when properly harnessed, can serve as useful
calibration to compensate and mitigate localization errors.

To exploit all this neighboring RSS information, we propose the
following COTAR localization strategy depicted in Figure
\ref{fig:scenario}:

\begin{figure}[htbp]
\centering
\includegraphics[width=3.8in]{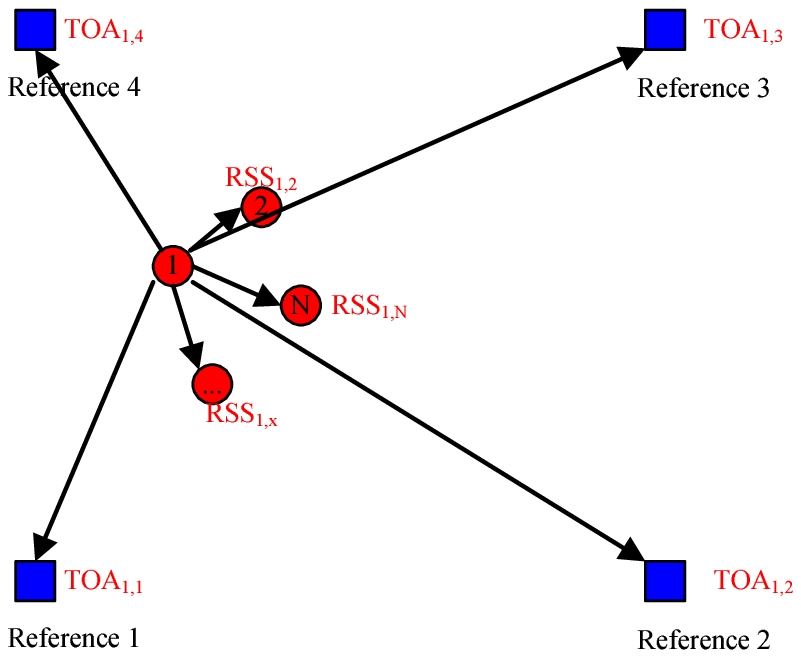}
\begin{center}(A)\end{center}
\includegraphics[width=3.8in]{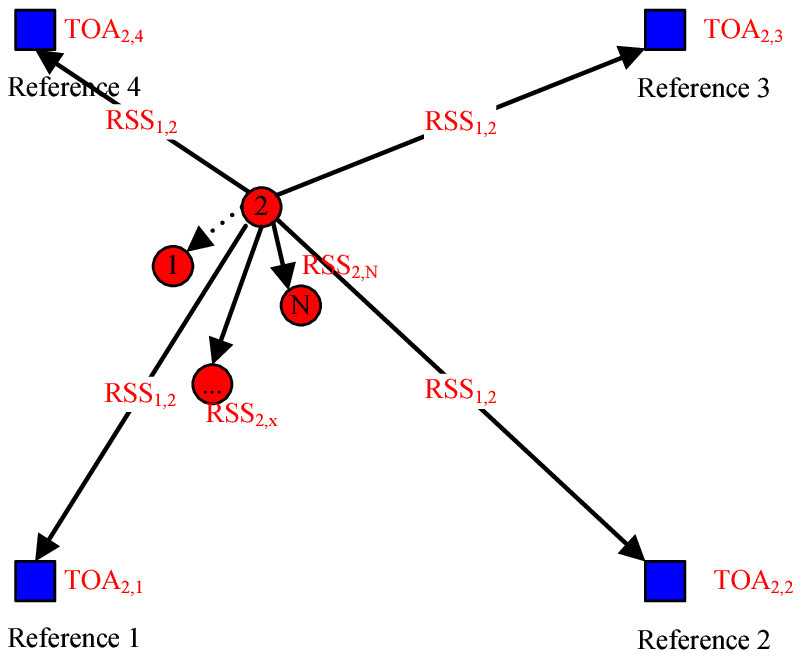}
\begin{center}(B)\end{center}
\vspace{-0.8cm}
\caption{System model for COTAR with $N$ cooperating
mobile nodes and $M=4$ reference nodes. (A) Step 1: Target node 1
broadcasts a beacon packet, all the other $(N\!-\!1)$ target nodes
measure the RSS (from node 1), and all the $M$ reference nodes
measure the TOA (from node 1). (B) Step 2: Target node 2 embeds the
RSS information it measured previously (i.e. RSS between node 1 and
node 2) in its beacon packet and broadcasts, all the subsequent
$(N\!-\!2)$ target nodes measure the RSS (from node 2), and all the
$M$ reference nodes measure the TOA (from node 2). In addition, the
$M$ reference nodes also try to decode the beacon packet to deduce the
embedded RSS information. Next, target node 3 will air an
RSS-bearing beacon packet. The procedure continues, until all the
$N$ collaborating target nodes have  sent a beacon packet. All the
$M$ reference nodes will combine all the TOA and RSS information and
perform a joint position estimation for all the $N$ target nodes.}
\label{fig:scenario}
\end{figure}

The first target node starts out sending a beacon signal, and all the
reference nodes measure and record the TOA of the beacon
packet. At the same time, all the subsequent target nodes also
measure the RSS from the transmitting target node.

Starting from the second target node, each target node will embed
all the RSS it has measured from the previous target nodes in the
payload of its beacon packet, and broadcasts. That is, the second
target node transmits to all the reference nodes the RSS information
between the first target node and itself; the third target node
transmits the RSS information between the first target node and
itself and between the second target node and itself; and so on. The
$M$ reference nodes not only measure and record the TOA of these
beacon packets, but also try to demodulate and decode the beacon
packets to obtain the  embedded RSS information.

Finally, the $M$ reference nodes share all the TOA information
(between targets nodes and reference nodes) and all the RSS
information (between target nodes) they have collected and perform
centralized computation, either at one of the reference nodes or a
dedicated control center. The centralized computation estimates the
positions of all the $N$ target nodes in a single batch, and may
implement different estimation methods, some of which are optimal
and others suboptimal. In this paper, we consider formulating the
problem as as a maximum likelihood position estimation problem, and
further propose a linear simplification of the ML method to render a
fast and efficient implementation through iterative refinement.

Here are a few comments:

(1) The proposed cooperative strategy is simple and
practical, and incurs very minimal cooperative overhead. Compared to
non-cooperative localization strategies where the target nodes must
each send a beacon packet anyway, the additional requirement imposed
by COTAR is for the target nodes to record and include  in its
beacon packet the RSS measurements. In practical systems, any
(wireless) packets, including load-less beacon packets, must consist
a sufficient packet-head to provide such information as where the
packet is originated from. In many systems, there is also a minimal
payload length required for a wireless packet, such that the packet
does not become too short to easily slip off without being detected.
As such, in which case the RSS information actually gets a free
ride.

(2) The proposed strategy works for any set of $M$ target nodes that
are in the vicinity of each other and willing to cooperate, but a
practical question arises as what is a good value for $N$. It is
expected that  the overall localization performance will improve
with $N$ (with a diminishing return), but  to engage many nodes in
cooperation can be expensive or difficult in practical scenarios.
Further, the communication overhead (recall that the $i$th target
node is expected to convey $(i-1)$ RSS measurements to the reference
nodes) as well as the complexity of the centralized estimation also
increase with $M$. As we will show later in our analytical and
simulation results, it is actually quite unnecessary to pull many
nodes. Two to four cooperating nodes are sufficient, and in many
cases, a mere engagement of two nodes ($N=2$) promises to delivery
most of the cooperative gain.

(3) In the COTAR strategy, the reference nodes are expected to not
only measure the TOA of beacon packets, but also extract from them
the embedded RSS information. It is not necessary for all the
reference nodes to correctly decode all the RSS information; as long
as one out of the $M$ reference nodes gets it (and conveys it to the
central node), an optimal joint estimation can proceed. In the rare
case when none of the reference nodes succeeds in deducing a
particular RSS information, the system may adopt one of the
following two strategies. (i) First, the $M$ reference nodes can
forward their individual reception of the beacon packets using, for
example, amplify-and-forward (AF) or other ostensible
signal-relaying technologies \cite{DAF_Bao}, to the central node for
joint decoding. The central node is thus supplied with $M$ noisy
copies of the beacon packet  with a diversity order of at least $M$,
and has a considerably higher probability of extracting the RSS
information than individual reference nodes can. (ii) Alternatively
or as a last resort, the system can simply give up on the missing
RSS information, and perform joint localization without having all
the RSS information in place. Depending on how many and what RSS
measurements are missing, the localization accuracy may be affected.
We will discuss this issue in detail later, but in general,
considering the rather small probability that none of the $M$
reference nodes has deduced the RSS information, the overall impact
of missing RSS is very small.

\section{Maximum Likelihood Position Estimation Using Joint TOA and
RSS}\label{sec:algorithm}
Having discussed the general COTAR protocol, we now elaborate the
maximum likelihood algorithm the central node performs to
simultaneously estimate all the target positions  through the
collection of TOA and RSS information.

\subsection{Path Loss among Target Nodes}

Let $\langle\mathbf{x},\mathbf{y}\rangle$ be the horizontal and vertical
coordinates of the actual positions for the $N$ targets nodes
($\mathbf{x}$ and $\mathbf{y}$ are each $N$ dimensional column
vectors). From the radio propagation attenuation model, we can write
the average path loss between a pair of target nodes $p$ and $q$, in
the unit of dB, as:
\begin{align} 
\label{eqn:TDOARSS_NEIGHBORING_RSS}
g_{p,q}=10\eta\log_{10}\!\left(\!\sqrt{(x^{(p)}-x^{(q)})^2-(y^{(p)}-y^{(q)})^2}\right)+g_0,
\ \ \ \ p,q\in [1..N], \mathbf{and\ } p\neq q,
\end{align}
where $g_0$ is the path-loss at the calibration distance, which is a
constant and does not affect the localization accuracy.

The effect of multi-path, scattering, and shadowing phenomena may be
collectively modeled as a multiplicative channel coefficient that
follows a logarithmically Gaussian distribution. When expressing the
channel attenuation in dB, the multiplicative coefficient becomes an
addictive (noise) term $\mathbf{Z_1}$ that follows the Gaussian distribution
(in dB). Since there are $N$ target nodes, the actual path loss
between any pair of target nodes can be expressed in  an ${N \choose
2}$ dimensional column vector as
\begin{align}
\mathbf{g}^*_{\left(N\atop 2\right)\times
1}=\mathbf{g}_{\left(N\atop 2\right)\times
1}+\mathbf{Z_1}_{\left(N\atop 2\right)\times 1}.
\end{align}

\subsection{Time-Of-Arrival from Target Nodes to Reference Nodes}

The distances between the $i$th reference node and all the $N$
target nodes can be expressed using an $M$ dimensional column vector function:
\begin{align}\label{eqn:TDOARSS_dist}
\mathbf{d}_{i}&=\sqrt{(\mathbf{x}_r-x^{(i)})^2+(\mathbf{y}_r-y^{(i)})^2}, \quad\ \  i=1\cdots N.
\end{align}

The time of arrival (TOA) for the $i$th target node's signal arrives at all the $M$ reference nodes will ideally take the form as follows:
\begin{align}\label{eqn:TDOARSS_dist_toa}
\mathbf{t}_{i}=\mathbf{d}_{i}/c, \ \ \ \ i\in[1..N],
\end{align}
where $c$ is the speed of the waveform propagation, which is $299792458$ m/s for radio frequency signals.
We arrange all TOAs from each target nodes to each reference nodes into one single $MN$ dimensional TOA vector $\mathbf{t}$ as
\begin{align}
\mathbf{t}&=[\mathbf{t}_1,\mathbf{t}_2,\cdots,\mathbf{t}_N]^T.
\end{align}

All this TOA estimation error is modeled as an $MN$ dimensional
Gaussian-distributed error vector $\mathbf{\tau}$. Hence, the noisy TOA vector from each
target node to each reference node can be written as:
\begin{align}
\mathbf{t}^*&=\mathbf{t}+\mathbf{\tau}.
\end{align}

\subsection{Path Loss between Target Nodes and Reference nodes}

The average path loss between each target node $i$ and all reference nodes, in
the unit of dB, as:
\begin{align}
\label{eqn:TDOARSS_remote_RSS}
\mathbf{h}_{i}=10\eta\log_{10}\!\left(\!\sqrt{(\mathbf{x}_r-x^{(i)})^2-(\mathbf{y}_r-y^{(i)})^2}\right)+g_0,
\ \ \ \ p,q\in [1..N], \mathbf{and\ } p\neq q,
\end{align}
where $g_0$ is the path-loss at the calibration distance, which is a
constant and does not affect the localization accuracy. We also rearrange all the RSS between each target node and each reference node in a single $MN$ dimensional column vector $\mathbf{h}$ as:
\begin{align}
\mathbf{h}&=[\mathbf{h}_1,\mathbf{h}_2,\cdots,\mathbf{h}_N]^T.
\end{align}

The effect of multi-path, scattering, and shadowing phenomena may be
collectively modeled as a multiplicative channel coefficient that
follows an addictive Gaussian noise term $\mathbf{Z_2}$ in logarithmical domain using unit dB.
Since there are $N$ target nodes, the actual path loss
between each target node and any reference node can be expressed in  an $M\times N$ dimensional column vector as
\begin{align}
\mathbf{h}^*_{\left(MN \right)\times
1}=\mathbf{h}_{\left(MN\right)\times
1}+\mathbf{Z_2}_{\left(MN\right)\times 1}.
\end{align}

\subsection{Unified RSS and TOA observation vector function}

For a compact representation, we subsume the expected path-loss
vector $\mathbf{g}$ and all the expected TOAs $\mathbf{t}_{i}$ in a single
column vector consisting of
${N\choose 2}+2MN$ rows, termed {\it observation vector}:
\begin{align}
\mathbf{f}(\mathbf{x},\mathbf{y})=(\overbrace{\mathbf{g}}^{N\choose 2},\overbrace{\mathbf{t}}^{MN},\overbrace{\mathbf{h}}^{MN})^T.
\end{align}

The true observation can be written as the sum of the observation vector
$\mathbf{f}(\mathbf{x},\mathbf{y})$ and a Gaussian noise vector
$\mathbf{n}$:
\begin{align}
\mathbf{r}=&\mathbf{f}(\mathbf{x},\mathbf{y})+\mathbf{n},\label{eqn:measurements}
\end{align}
where the noise vector $\mathbf{n}$ is given by
\begin{align}
\mathbf{n}=(\mathbf{Z_1},\mathbf{\tau},\mathbf{Z_2})^T.
\end{align}

The observation error $\mathbf{n}$ denotes the measurement
distortion caused by various imperfect conditions regarding RSS and
TOA, and is assumed to be a multivariate random vector with
$D={N\choose 2}+2MN$ dimensional positive-definite covariance
matrix:
\begin{align}
\mathbf{\Lambda}&=E\left[(\mathbf{n}-E[\mathbf{n}])(\mathbf{n}-E[\mathbf{n}])^T\right]\nonumber \\
&=\rm{diag}\Big( \overbrace{\sigma_{Z_1}^2}^{\left(N\atop 2\right)}, \overbrace{{\bf \sigma}_{\tau}^2}^{MN},\overbrace{{\bf \sigma}_{Z_2}^2}^{MN}
\Big).
\end{align}

\subsection{Simplification through Taylor Expansion}

The vector function $\mathbf{g}(\mathbf{x},\mathbf{y})$,
$\mathbf{t}(\mathbf{x},\mathbf{y})$, and $\mathbf{h}(\mathbf{x},\mathbf{y})$ are nonlinear, which makes
the measurement vector function
$\mathbf{f}(\mathbf{x},\mathbf{y})$ technically challenging to
handle. To make real-time computation possible, we approximate them
using a single linear function by Taylor expanding them at the
vector point $\langle\, \mathbf{x}_0,\mathbf{y}_0\,\rangle$.
Mathematically, it can be written as follows:
\begin{align}\label{eqn:Taylar_expansion}
\mathbf{f}(\mathbf{x},\mathbf{y})\approx
\mathbf{f}(\mathbf{x}_0,\mathbf{y}_0)+
{\bf G}\times \left(\mathbf{x}-\mathbf{x}_0\atop \mathbf{y}-\mathbf{y}_0\right)^T,
\end{align}
where ${\bf G}$ is the partial derivative matrix for each
measurement on each unknown variable. The partial derivative matrix
can be further written a $D\times 2N$ dimensional matrix as follows:
\begin{align}
{\bf G}=[G_{i,j}]_{D\times 2N}=\left[ \begin{array}{cccccc}
      \frac{\partial f_1}{\partial x_{1}}\big|_{\mathbf{x}=\mathbf{x}_0\atop \mathbf{y}=\mathbf{y}_0}&   \cdots&\frac{\partial f_1}{\partial x_{N}}\big|_{\mathbf{x}=\mathbf{x}_0\atop \mathbf{y}=\mathbf{y}_0}&      \frac{\partial f_1}{\partial y_{1}}\big|_{\mathbf{x}=\mathbf{x}_0\atop \mathbf{y}=\mathbf{y}_0}&\cdots&\frac{\partial f_1}{\partial y_{N}}\big|_{\mathbf{x}=\mathbf{x}_0\atop \mathbf{y}=\mathbf{y}_0}\\
      \frac{\partial f_2}{\partial x_{1}}\big|_{\mathbf{x}=\mathbf{x}_0\atop \mathbf{y}=\mathbf{y}_0}&   \cdots&\frac{\partial f_2}{\partial x_{N}}\big|_{\mathbf{x}=\mathbf{x}_0\atop \mathbf{y}=\mathbf{y}_0}&      \frac{\partial f_2}{\partial y_{1}}\big|_{\mathbf{x}=\mathbf{x}_0\atop \mathbf{y}=\mathbf{y}_0}&\cdots&\frac{\partial f_2}{\partial y_{N}}\big|_{\mathbf{x}=\mathbf{x}_0\atop \mathbf{y}=\mathbf{y}_0}\\
      \vdots                                                                                                  &         &                 \vdots                                                                                    &                         \vdots                                                                              &      & \vdots            \\
      \frac{\partial f_D}{\partial x_{1}}\big|_{\mathbf{x}=\mathbf{x}_0\atop \mathbf{y}=\mathbf{y}_0}&   \cdots&\frac{\partial f_D}{\partial x_{N}}\big|_{\mathbf{x}=\mathbf{x}_0\atop \mathbf{y}=\mathbf{y}_0}&      \frac{\partial f_D}{\partial y_{1}}\big|_{\mathbf{x}=\mathbf{x}_0\atop \mathbf{y}=\mathbf{y}_0}&\cdots&\frac{\partial f_D}{\partial y_{N}}\big|_{\mathbf{x}=\mathbf{x}_0\atop \mathbf{y}=\mathbf{y}_0}\\
    \end{array}\right],
\end{align}
whose elements $G_{i,j \in [1, N]}$ and $G_{i,j\in [N+1,2N]}$
represent the partial derivatives of the $i$th element in the
observation vector with respect to the $j$th target node's
horizontal coordinate $x^{(j)}$ and vertical coordinate
$y^{(j)}$ respectively. $D={N \choose 2} +2MN$ is the dimension of
the observation vector $\mathbf{f}$, and $\langle \mathbf{x}_0,\mathbf{y}_0\rangle$ is the initial position vector.

\subsection{Partial Derivative Matrix ${\bf G}$}

A key computation step in the proposed joint estimation is the
computation of the partial derivative matrix ${\bf G}$, which
consists of two parts:
\begin{align}
{\bf G}=\left(
              \begin{array}{c}
               {\bf  A}   \\
               {\bf B}    \\
               {\bf C}    \\
              \end{array}
            \right),
\end{align}
where the sub-matrices ${\bf A}$ is the partial
derivative matrix pertaining to the RSS measurements between any two target
nodes, ${\bf B}$ is the TOA measurements between target nodes and
reference nodes), and $\mathbf{C}$ is the RSS measurement between each target node and each reference node, respectively.

The sub-matrix ${\bf A}$ has ${ N \choose 2}$ rows and $2N$ columns,
where rows correspond to pairs of target nodes, and columns
correspond to the target nodes' horizontal and vertical positions.
The exact formulation of ${\bf A}$ is
defined as follows:
\begin{align}\label{eqn:RSS_A}
{\bf A}=\left(
    \begin{array}{c}
      {\bf A}_{1,2}\\
        {\bf A}_{1,3}\\
         \cdots \\
          {\bf A}_{N-1,N}
    \end{array}
  \right). \ \ \ \ \ \ \ \ (p\neq q)
\end{align}

A row vector ${\bf A}_{p,g}$ corresponds to the RSS between the
$p$th and $q$th target nodes. It has four non-zero elements in the
positions $p$, $q$, $(N\!+\!p)$ and $(N\!+\!q)$, and is zero
everywhere else:
\begin{align}\label{eqn:RSS_A_pq}
{\bf A}_{p,q}&=\left(
          \begin{array}{c}
            \cdots0,
  \underbrace{\frac{\partial g}{\partial x^{(p)}}
   \Big|_{\mathbf{x}=\mathbf{x}_0\atop \mathbf{y} =\mathbf{y}_0}}_{p\mbox{th}}, 0\cdots0,
  \underbrace{\frac{\partial g}{\partial x^{(q)}}
  \Big|_{\mathbf{x}=\mathbf{x}_0\atop \mathbf{y}=\mathbf{y}_0}}_{q\mbox{th}},
   0\cdots0,
   \underbrace{\frac{\partial g}{\partial y^{(p)}}
   \Big|_{\mathbf{x}=\mathbf{x}_0\atop \mathbf{y}=\mathbf{y}_0}}_{(N+p)\mbox{th}}, 0\cdots0,
    \underbrace{\frac{\partial g}{\partial y^{(q)}}
   \Big|_{\mathbf{x}=\mathbf{x}_0\atop \mathbf{y}=\mathbf{y}_0}}_{(N+q)\mbox{th}}, 0\cdots
         \end{array}
        \right)
\end{align}

Plugging (\ref{eqn:TDOARSS_NEIGHBORING_RSS}) in (\ref{eqn:RSS_A_pq}) and after
simplification, we obtain the following:
\begin{align}
\frac{\partial g}{\partial x^{(p)}}\Big|_{\mathbf{x}=\mathbf{x}_0\atop \mathbf{y} =\mathbf{y}_0}=-\frac{\partial g}{\partial x^{(q)}}\Big|_{\mathbf{x}=\mathbf{x}_0\atop \mathbf{y}=\mathbf{y}_0}=
\frac{10\eta\ln(10)(x_0^{(p)}-x_0^{(q)})}{(x_0^{(p)}-x_0^{(q)})^2\!+(y_0^{(p)}-y_0^{(q)})^2},\\
\frac{\partial g}{\partial y^{(p)}}\Big|_{\mathbf{x}=\mathbf{x}_0\atop \mathbf{y} =\mathbf{y}_0}=-\frac{\partial g}{\partial y^{(q)}}\Big|_{\mathbf{x}=\mathbf{x}_0\atop \mathbf{y}=\mathbf{y}_0}=
\frac{10\eta\ln(10)(y_0^{(p)}-y_0^{(q)})}{(x_0^{(p)}-x_0^{(q)})^2\!+(y_0^{(p)}-y_0^{(q)})^2}.
\end{align}

The partial derivative matrix for TOA, ${\bf B}$, can be further
decomposed to $2N$ sub-matrices, each corresponding to the TOA
derivative matrix for one target nodes with respect to its
horizontal and vertical coordinates:
\begin{align}
{\bf B}=\left(
    \begin{array}{cccccccc}
      {\bf B}_x^{(1)} &                 &       &                &{\bf B}_y^{(1)}&               &      &                   \\
                      & {\bf B}_x^{(2)} &       &                &               &{\bf B}_y^{(2)}&      &                   \\
                      &                 &\cdots &                &               &               &\cdots&                   \\
                      &                 &       & {\bf B}_x^{(N)}&               &               &      &{\bf B}_y^{(N)}    \\
    \end{array}
  \right)
\end{align}

Each submatrix ${\bf B}_x^{(i)}$ is an $M$ dimensional column
matrix, denoting the partial derivatives of the $M$ TOA values,
measured by the $M$ reference nodes about the $i$th target node,
with respect to the $i$th target node's horizontal coordinates:
\begin{align}
{\bf B}_x^{(i)}&=\left(
      \left.\frac{\partial \mathbf{t}_i^{(1)}}{\partial x^{(i)}}\right|_{x^{(i)}=x^{(i)}_0\atop y^{(i)}=y^{(i)}_0},\
      \left.\frac{\partial \mathbf{t}_i^{(2)}}{\partial x^{(i)}}\right|_{x^{(i)}=x^{(i)}_0\atop y^{(i)}=y^{(i)}_0},\
      \cdots,\
      \left.\frac{\partial \mathbf{t}_i^{(M)}}{\partial x^{(i)}}\right|_{x^{(i)}=x^{(i)}_0\atop y^{(i)}=y^{(i)}_0}
  \right)^T.
\end{align}

Submatrices ${\bf B}_y^{(i)}$ are similar, but the partial
derivative is with respect to the the vertical coordinate pf the
$i$th target node:
\begin{align}
{\bf B}_y^{(i)}&=\left(
      \left.\frac{\partial \mathbf{t}^{(1)}_i}{\partial y^{(i)}}\right|_{x^{(i)}=x^{(i)}_0\atop y^{(i)}=y^{(i)}_0},\
      \left.\frac{\partial \mathbf{t}^{(2)}_i}{\partial y^{(i)}}\right|_{x^{(i)}=x^{(i)}_0\atop y^{(i)}=y^{(i)}_0},\
      \cdots,\
      \left.\frac{\partial \mathbf{t}^{(M)}_i}{\partial y^{(i)}}\right|_{x^{(i)}=x^{(i)}_0\atop y^{(i)}=y^{(i)}_0}
  \right)^T.
\end{align}

Combining (\ref{eqn:TDOARSS_dist}) and (\ref{eqn:TDOARSS_dist_toa}),
and after simplification, the partial derivative sub-matrices can be written as follows:
\begin{align}
{\bf B}_x^{(i)}&=\frac{1}{c}\left(
    \begin{array}{cccc}
    \frac{x_r^{(1)}-x_0^{(i)}}{\sqrt{(x_r^{(1)}-x_0^{(i)})^{2} + (y_r^{(1)}-y_0^{(i)})^{2}}}\\
    \frac{x_r^{(2)}-x_0^{(i)}}{\sqrt{(x_r^{(2)}-x_0^{(i)})^{2} + (y_r^{(2)}-y_0^{(i)})^{2}}}\\
        \cdots\\
    \frac{x_r^{(M)}-x_0^{(i)}}{\sqrt{(x_r^{(M)}-x_0^{(i)})^{2} + (y_r^{(M)}-y_0^{(i)})^{2}}}
    \end{array}
    \right)_{M\times 1},\ \ \ \ i\in[1..N]
\end{align}

\begin{align}
{\bf B}_y^{(i)}=\frac{1}{c}\left(
    \begin{array}{cccc}
        \frac{y_r^{(1)}-y_0^{(i)}}{\sqrt{(x_r^{(1)}-x_0^{(i)})^{2} + (y_r^{(1)}-y_0^{(i)})^{2}}}\\
        \frac{y_r^{(2)}-y_0^{(i)}}{\sqrt{(x_r^{(2)}-x_0^{(i)})^{2} + (y_r^{(2)}-y_0^{(i)})^{2}}}\\
        \cdots\\
        \frac{y_r^{(M)}-y_0^{(i)}}{\sqrt{(x_r^{(M)}-x_0^{(i)})^{2} + (y_r^{(M)}-y_0^{(i)})^{2}}}
    \end{array}
  \right)_{M\times 1},\ \ \ \ i\in[1..N]
\end{align}
where $c$ is the speed of light.

The partial derivative matrix for RSS between each target node and each reference node, ${\bf C}$, can be further
decomposed to $2N$ sub-matrices, each corresponding to the TOA
derivative matrix for one target nodes with respect to its
horizontal and vertical coordinates:
\begin{align}
{\bf C}=\left(
    \begin{array}{cccccccc}
      {\bf C}_x^{(1)} &                 &       &                &{\bf C}_y^{(1)}&               &      &                   \\
                      & {\bf C}_x^{(2)} &       &                &               &{\bf C}_y^{(2)}&      &                   \\
                      &                 &\cdots &                &               &               &\cdots&                   \\
                      &                 &       & {\bf C}_x^{(N)}&               &               &      &{\bf C}_y^{(N)}    \\
    \end{array}
  \right)
\end{align}

Each submatrix ${\bf C}_x^{(i)}$ is an $M$ dimensional column
matrix, denoting the partial derivatives of the $M$ RSS values,
measured by the $M$ reference nodes about the $i$th target node,
with respect to the $i$th target node's horizontal coordinates:
\begin{align}
{\bf C}_x^{(i)}&=\left(
      \left.\frac{\partial \mathbf{h}^{(1)}_i}{\partial x^{(i)}}\right|_{x^{(i)}=x^{(i)}_0\atop y^{(i)}=y^{(i)}_0},\
      \left.\frac{\partial \mathbf{h}^{(2)}_i}{\partial x^{(i)}}\right|_{x^{(i)}=x^{(i)}_0\atop y^{(i)}=y^{(i)}_0},\
      \cdots,\
      \left.\frac{\partial \mathbf{h}^{(M)}_i}{\partial x^{(i)}}\right|_{x^{(i)}=x^{(i)}_0\atop y^{(i)}=y^{(i)}_0}
  \right)^T.
\end{align}

Submatrices ${\bf C}_y^{(i)}$ are similar, but the partial
derivative is with respect to the the vertical coordinate pf the
$i$th target node:
\begin{align}
{\bf C}_y^{(i)}&=\left(
      \left.\frac{\partial \mathbf{h}^{(1)}_i}{\partial y^{(i)}}\right|_{x^{(i)}=x^{(i)}_0\atop y^{(i)}=y^{(i)}_0},\
      \left.\frac{\partial \mathbf{h}^{(2)}_i}{\partial y^{(i)}}\right|_{x^{(i)}=x^{(i)}_0\atop y^{(i)}=y^{(i)}_0},\
      \cdots,\
      \left.\frac{\partial \mathbf{h}^{(M)}_i}{\partial y^{(i)}}\right|_{x^{(i)}=x^{(i)}_0\atop y^{(i)}=y^{(i)}_0}
  \right)^T.
\end{align}

Combining (\ref{eqn:TDOARSS_dist}) and (\ref{eqn:TDOARSS_dist_toa}),
and after simplification, the partial derivative sub-matrices can be written as follows:
\begin{align}
{\bf C}_x^{(i)}&=\alpha\!\left(
    \begin{array}{cccc}
    \frac{x_r^{(1)}-x_0^{(i)}}{(x_r^{(1)}-x_0^{(i)})^{2} + (y_r^{(1)}-y_0^{(i)})^{2}}\\
    \frac{x_r^{(2)}-x_0^{(i)}}{(x_r^{(2)}-x_0^{(i)})^{2} + (y_r^{(2)}-y_0^{(i)})^{2}}\\
        \cdots\\
    \frac{x_r^{(M)}-x_0^{(i)}}{(x_r^{(M)}-x_0^{(i)})^{2} + (y_r^{(M)}-y_0^{(i)})^{2}}
    \end{array}
    \right)_{M\times 1},{\bf C}_y^{(i)}=\alpha\!\left(
    \begin{array}{cccc}
        \frac{y_r^{(1)}-y_0^{(i)}}{(x_r^{(1)}-x_0^{(i)})^{2} + (y_r^{(1)}-y_0^{(i)})^{2}}\\
        \frac{y_r^{(2)}-y_0^{(i)}}{(x_r^{(2)}-x_0^{(i)})^{2} + (y_r^{(2)}-y_0^{(i)})^{2}}\\
        \cdots\\
        \frac{y_r^{(M)}-y_0^{(i)}}{(x_r^{(M)}-x_0^{(i)})^{2} + (y_r^{(M)}-y_0^{(i)})^{2}}
    \end{array}
  \right)_{M\times 1},\ \ \ \ i\in[1..N]
\end{align}
, where $\alpha$ is the constant of $10\eta\ln(10)$, and $\eta=3.086$ is the channel attenuation factor.

\subsection{Joint ML Estimation with Linear Approximation }

Having computed the partial derivative matrix ${\bf G}$, we now
discuss position estimation through maximum likelihood method and
its efficient implementation.

Since the noise in the observation vector function
$\mathbf{r}(\mathbf{x}, \mathbf{y})$ is assumed Gaussian distributed, we can
write the probability distribution function (pdf) at the vector point
$\langle\, \mathbf{x}_0,\mathbf{y}_0\,\rangle$ as:
\begin{align}
&p\!\left(\!\mathbf{r}\big|_{{\bf x}={\bf x}_0 \atop {\bf y}={\bf y}_0}\!\right)=
\frac{1}{(2\pi)^{D/2}\!|\mathbf{\Lambda}|^{1/2}}\!\exp\left\{\frac{-1}{2}[\mathbf{r}-\mathbf{f}(\mathbf{x}_0,\mathbf{y}_0)]^T\mathbf{\Lambda}^{-1}[\mathbf{r}-\mathbf{f}(\mathbf{x}_0,\mathbf{y}_0)]\right\},
\label{eqn:pdf}
\end{align}
where $|\mathbf{\Lambda}|$ denotes the determinant of $\mathbf{\Lambda}$, the
superscript $-1$ denotes the matrix inverse, and $D={N\choose 2}+MN$ is the dimension of $\mathbf{\Lambda}$.

The maximum likelihood estimator calculates the value
$\langle\, \mathbf{x},\mathbf{y}\,\rangle$ that maximizes (\ref{eqn:pdf}). Equivalently, it
minimizes the quadratic term:
\begin{align}
Q(\mathbf{x},\mathbf{y})&=[\mathbf{r}-\mathbf{f}(\mathbf{x},\mathbf{y})]^T\mathbf{\Lambda}^{-1}[\mathbf{r}-\mathbf{f}(\mathbf{x},\mathbf{y})].
\label{eqn:Quadratic}
\end{align}

We further define
\begin{align}\label{eqn:r1}
\mathbf{r}_1&=\mathbf{r}-\mathbf{f}(\mathbf{x}_0,\mathbf{y}_0)+\bf{G}\left(\mathbf{x}_0 \atop \mathbf{y}_0\right).
\end{align}

Plugging (\ref{eqn:r1}) into (\ref{eqn:Quadratic}), the quadratic term $Q(\mathbf{x},\mathbf{y})$ can be further written as:
\begin{align}
Q(\mathbf{x},\mathbf{y})&=\left[\mathbf{r}_1-\bf{G}\left(\mathbf{x}\atop \mathbf{y}\right)\right]^T\mathbf{\Lambda}^{-1}\left[\mathbf{r}_1-\bf{G}\left(\mathbf{x}\atop \mathbf{y}\right)^T\right].
\end{align}

Minimizing $Q(\mathbf{x},\mathbf{y})$ is a reasonable criterion
for determination of an estimator even when the measurement error
cannot be assumed to be Gaussian distributed. In this case, the
resulting estimator is called the least squares estimator and
$\mathbf{\Lambda}^{-1}$ is regarded as a matrix of weighting
coefficients. Therefore, we turn the position estimation problem to
solving the equation of $\partial Q(\mathbf{x},\mathbf{y})/
\partial \mathbf{x}\partial \mathbf{y} = \mathbf{0}$.

From its definition, $\mathbf{\Lambda}$ is a symmetric matrix, i.e.,
$\mathbf{\Lambda}=\mathbf{\Lambda}^T$. Since
$\left(\mathbf{\Lambda}^{-1}\right)^T=\left(\mathbf{\Lambda}^T\right)^{-1}$, it
follows that $\left(\mathbf{\Lambda}^{-1}\right)^T=\mathbf{\Lambda}^{-1}$, which shows
that $\mathbf{\Lambda}^{-1}$ is a symmetric matrix. Therefore, we can obtain:
\begin{align}\label{eqn:drivative_equal_0}
\partial Q(\mathbf{x},\mathbf{y})/
\partial \mathbf{x}\partial \mathbf{y} =2\mathbf{G}^T\mathbf{\Lambda}^{-1}\mathbf{G}\left(\hat{\mathbf{x}}\atop \hat{\mathbf{y}}\right)-2\mathbf{G}^T\mathbf{\Lambda}^{-1}\mathbf{r}_1= \mathbf{0}.
\end{align}

We assume that the matrix of $\mathbf{G}^T\mathbf{\Lambda}^{-1}\mathbf{G}$ is
nonsingular. Thus the solution of (\ref{eqn:drivative_equal_0}) can
be expressed as:
\begin{align}\label{eqn:solution1}
\left(\hat{\mathbf{x}}\atop\hat{\mathbf{y}}\right)&=\left(\mathbf{G}^T\mathbf{\Lambda}^{-1}\mathbf{G}\right)^{-1}\mathbf{G}^T\mathbf{\Lambda}^{-1}\mathbf{r}_1.
\end{align}

Plugging (\ref{eqn:r1}) in (\ref{eqn:solution1}), and after simplification, we arrive at the
following estimate for the target nodes' position vector:
\begin{align}
\left( \mathbf{\hat{x}}\atop \mathbf{\hat{y}} \right)
= \left( \mathbf{\hat{x}}\atop \mathbf{\hat{y}} \right)+\left({\bf G}^T\mathbf{\Lambda}^{-1}{\bf G}\right)^{-1}{\bf G}^T\mathbf{\Lambda}^{-1}\left(\mathbf{r}-\mathbf{f}(\mathbf{x}_0,\mathbf{y}_0)\right),
\label{eqn:estimator}
\end{align}
where the first term denotes the initial positions of these target
nodes, and the second term provides an incremental adjustment.

\subsection{Iterative Estimation Refinement}
Note that (\ref{eqn:estimator}) suffers from the inaccuracy caused
by approximating a nonlinear function
$\mathbf{f}(\mathbf{x},\mathbf{y})$ using the linear Taylor
expansion in (\ref{eqn:Taylar_expansion}) especially when $\langle\mathbf{x}_0,\mathbf{y}_0\rangle$
are far away from the actually position $\langle\mathbf{x},\mathbf{y}\rangle$. To mitigate the error and
improve the estimation accuracy in (\ref{eqn:estimator}), we propose
to solve this problem through iterative refinement, in which the
estimated position of the previous iteration
$\langle\mathbf{\hat{x}},\mathbf{\hat{y}}\rangle$ is used as
the initial position $\langle\mathbf{x}_0,\mathbf{y}_0\rangle$
for the next refining iteration. To start, when we have no {\it a
priori} knowledge, we may take the center point (or any point) of
the scenario as the initial location for all the target nodes. For
example, suppose $M=4$ reference nodes are placed in the corner of a
square area with $L$ meters per side, the proposed iterative
localization proceeds as follows:
\begin{equation}\label{eq:iterative_estimation}
\left\{\begin{aligned}
& \!\!\left(\! {\mathbf{\hat{x}}[1]}\atop{\mathbf{\hat{y}}[1]} \!\right)
=\left(L/2,\cdots,L/2\right)^T_{2N\times 1}, & \\
& \!\!\left(\! {\mathbf{\hat{x}}[k]}\atop{\mathbf{\hat{y}}[k]}
\!\right) \!=\!\left( { \mathbf{\hat{x}}[k\!-\!1] }\atop{
\mathbf{\hat{y}}[k\!-\!1] }\right)\!+\!\left(\mathbf{G}^T\!\mathbf{\Lambda}^{-1}\!\mathbf{G}\!\right)^{-1}\!\!\mathbf{G}^T\!\mathbf{\Lambda}^{-1}\!(\mathbf{r}\!-\!\mathbf{f}(\mathbf{\hat{x}}[k\!-\!1] ,\mathbf{\hat{y}}[k\!-\!1])\!),&\ k\!=\!2,3,\cdots
\end{aligned}\right.
\end{equation}
where $[k]$ is the iteration index.

It is expected that a good choice for the initial position
$\langle\mathbf{x}_0,\mathbf{y}_0\rangle$ helps expedite the
process while a bad one may delay the convergence, but this appears
to be a non-issue, since our algorithm converges very fast. Through
extensive simulations, we show that it generally takes no more than
two to three iterations to arrive at an accurate estimation, even
when the initial position is set very far from the actual position.

In the case of mobile tracking, when the previous-time position is
used as the initial position to estimate the next-time position, a
single iteration suffices. The iterative process not only improves
the localization accuracy, but also makes the strategy particularly
suitable for mobile localization. When target nodes are moving and
their continuous trace needs to be detected, it is very natural to
use their previous locations as the starting point to estimate the
next locations. Unless the nodes are moving at extremely fast
speeds, the new locations will not be far from the old ones, and
hence a single iteration suffices, allowing efficient position
tracking.

\section{Analysis of Estimation Accuracy and performance comparisons}
\label{sec:analysis}
To provide a theoretical support, we first evaluate the Cramer-Rao lower bound
for the proposed cooperative TOA-RSS localization scheme. We next analyze the theoretical root mean square error of the proposed maximum likelihood position estimation algorithm. We compare the bounds of the proposed localization scheme with the existing localization schemes.

\subsection{Cramer Rao lower bounds (CRB) analysis}
The CRB of an unbiased estimator $\langle\mathbf{\hat{x}},\mathbf{\hat{y}}\rangle$ is given by
$$\mathrm{cov}(\mathbf{\hat{x}},\mathbf{\hat{y}})\geq I(\mathbf{x},\mathbf{y})^{-1},$$
 where $I(.)$ is the Fisher information matrix (FIM) given by \cite{TDOA-RSS-Cramer-Rao}:
\begin{align}\label{eqn:CRB_FIM}
I(\mathbf{\hat{x}},\mathbf{\hat{y}})&=-E\nabla_{\mathbf{x},\mathbf{y}}\left(\nabla_{\mathbf{x},\mathbf{y}}\xi(\mathbf{r}|\mathbf{\mathbf{x,y,x_r,y_r}})\right),\nonumber\\
&=\left[
            \begin{array}{cc}
            I_{xx} & I_{xy} \\
            I_{xy} & I_{yy} \\
            \end{array}
\right],
\end{align}
where $\xi(\mathbf{r}|\mathbf{x,y,\mathbf{x_r,y_r}})$ is the logarithm of the joint conditional probability density function. We can further write
\begin{align}\label{eqn:CRB_FIM_sub}
I_{xx}&=-E\left[\frac{\partial^2\xi(\mathbf{r}|x,y,\mathbf{x_r,y_r})}{\partial x^2}\right],\nonumber\\
I_{xy}&=-E\left[\frac{\partial^2\xi(\mathbf{r}|x,y,\mathbf{x_r,y_r})}{\partial x \partial y}\right],\nonumber\\
I_{yy}&=-E\left[\frac{\partial^2\xi(\mathbf{r}|x,y,\mathbf{x_r,y_r})}{\partial y^2}\right].
\end{align}

The CRB on the standard derivation of each target node's position estimation is:
\begin{align}
\sigma_{CRB}&=\sqrt{\min_{\hat{x},\hat{y}} E\left[(\hat{x}-x)^2+(\hat{y}-y)^2\right]}=\sqrt{\min \mathrm{tr}\left[\mathrm{cov}(\hat{x},\hat{y})\right]}\nonumber\\
&=\sqrt{\mathrm{tr}\left[I(x,y)^{-1}\right]}\nonumber\\
&=\sqrt{\frac{I_{xx}+I_{yy}}{I_{xx}I_{yy}-I_{xy}^2}}.
\end{align}

The expression of $\xi(\mathbf{r}|x,y,\mathbf{x_r,y_r})$ is different for each localization estimation scheme.

\underline{Existing Localization Schemes}

The traditional RSS scheme \cite{Indoor}\cite{WSN_localization} uses only the RSS from a sensor node to all the $M$ reference nodes ($M$ RSS measures per sensor) to perform position estimation. We have:
\begin{align}\label{eqn:xi_RSS}
\xi_{RSS}(\mathbf{r}|x,y,\mathbf{x}_r,\mathbf{y}_r )=\sum_{i=1}^{M}\log p_{g_i|x,y,x_{ri},y_{ri}},
\end{align}
where $p_{g_i|x,y,x_{ri},y_{ri}}\sim \mathcal{N}\left(g_i (\mathrm{dB}), \sigma_z^2\right)$.

Similarly, for the traditional TOA scheme \cite{Indoor}\cite{WSN_localization}, only the TOA at the reference nodes $M$ TOA measures per sensor is used for estimation. The logarithm of the joint conditional probability density function can be written as:
\begin{align}\label{eqn:xi_TOA}
\xi_{TOA}(\mathbf{r}|x,y,\mathbf{x}_r,\mathbf{y}_r)&=\sum_{i=1}^{M}\log p_{t_i|x,y,x_{ri},y_{ri}},
\end{align}
where $p_{t_i|x,y,x_{ri},y_{ri}}\!\sim\! \mathcal{N}\left(t_i, \sigma_{\tau}^2\right)$, and $p_{t_i|x,y,x_{ri},y_{ri}}\!\sim\! \mathcal{N}\left(t_i, \sigma_{\tau}^2\right)$.

The existing non-cooperative hybrid TOA-RSS schemes exploit both the RSS and TOA from each senor to all the $M$ reference nodes (, which exploits total $2M$ modality measures per sensor, including $M$ modality measures per sensor for TOA and another $M$ modality measures per sensor for TOA). As discussed in \cite{TDOA-RSS-Cramer-Rao}, for the hybrid TOA-RSS scheme, it is given by
\begin{align}\label{eqn:xi_TOARSS}
&\xi_{TOA-RSS}(\mathbf{r}|x,y,\mathbf{x}_r,\mathbf{y}_r)\nonumber\\
=&\sum_{i=1}^{M}\log p_{g_i|x,y,x_{ri},y_{ri}}+\sum_{i=1}^{M}\log p_{t_i|x,y,x_{ri},y_{ri}}.
\end{align}

We further substitute (\ref{eqn:xi_RSS})(\ref{eqn:xi_TOA})(\ref{eqn:xi_TOARSS}) into (\ref{eqn:CRB_FIM})(\ref{eqn:CRB_FIM_sub}) respectively, and after some rather involved numerical calculation, we obtain the CRB for traditional RSS, TOA, and hybrid TOA-RSS localization schemes.

\underline{The Proposed New Scheme}

The new cooperative TOA-RSS scheme developed here makes use of not only the TOA from each sensor to all the $M$ reference nodes but also the RSS between the two neighboring sensors, to jointly estimate the poses of both sensors ( $2MN\!+\! {N\choose 2}$ modality measures for $N$ sensors). We can write the logarithm of the joint conditional probability density function as follows:
\begin{align}
\label{eqn:xi_COTAR}
\!\xi_{new}
(\!\mathbf{r}|\mathbf{x,y,x_r,y_r} \!)\!=\!\!\sum_{i=1}^{M}\!\log p_{t_i|x_p,y_p,x_{ri},y_{ri}}\!\!+\sum_{i=1}^{M}\log p_{t_i|x,y,x_{ri},y_{ri}}+\!\sum_{q=1,q\neq p}^N\log p_{g_{p,q}|\mathbf{x},\mathbf{y}}\!,
\end{align}
where $\log p_{g_{p,q}|\mathbf{x},\mathbf{y}}\sim \mathcal{N}\left(g_{p,q}, \sigma_z^2\right)$.

Similar to the procedure of calculation traditional localization schemes, substituting (\ref{eqn:xi_COTAR}) in (\ref{eqn:CRB_FIM}) and (\ref{eqn:CRB_FIM_sub}) and after simplification, we can compute the Cramer-Rao bound for the proposed new scheme.

\begin{figure*}[htbp]
\centering
\includegraphics[width=3in]{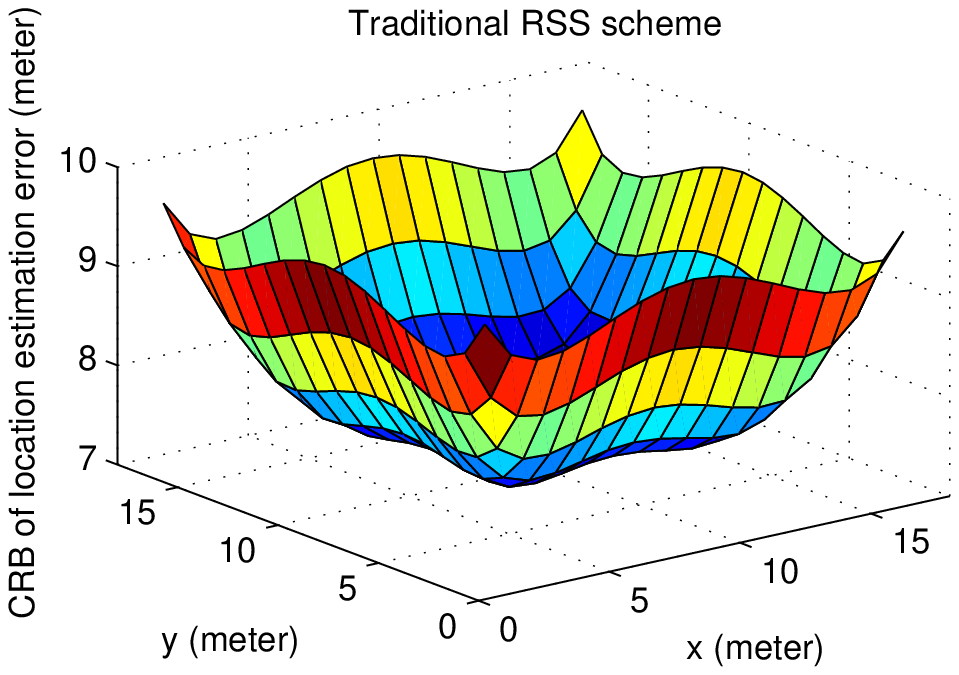}
\includegraphics[width=3in]{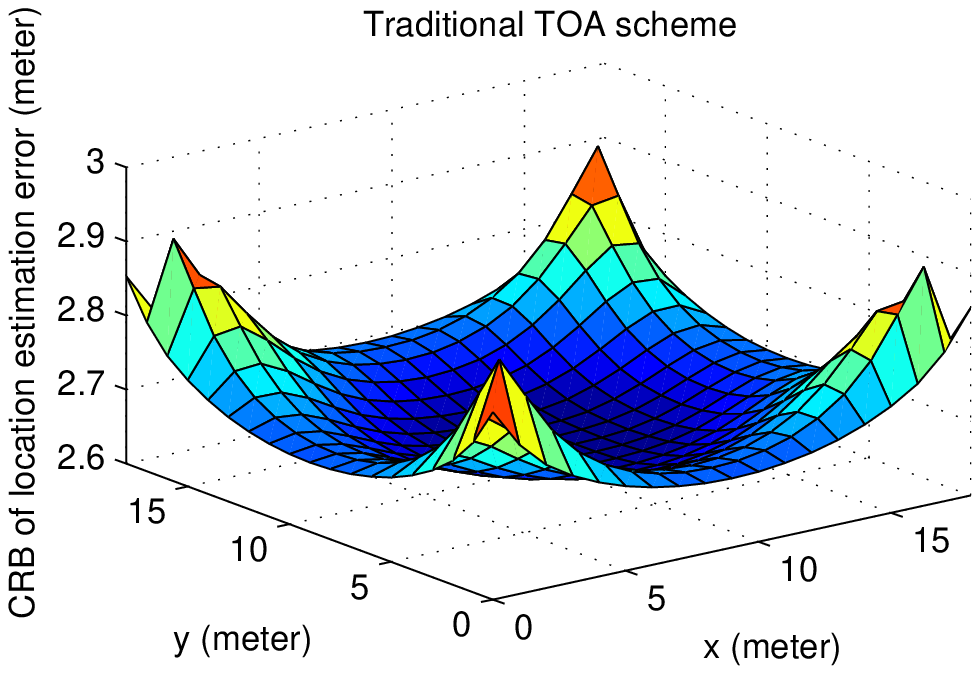}
\includegraphics[width=3in]{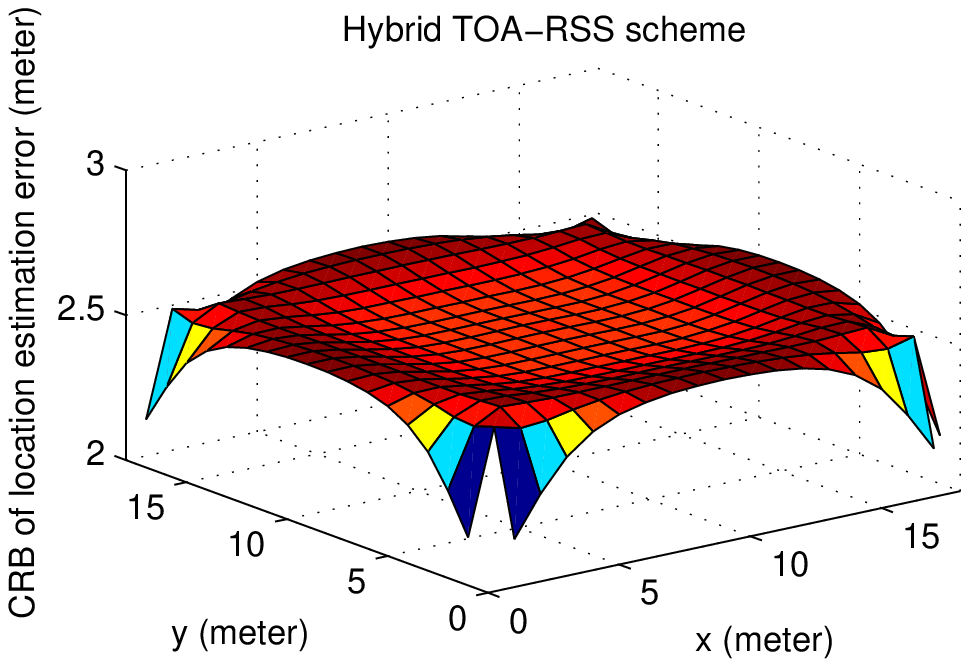}
\includegraphics[width=3in]{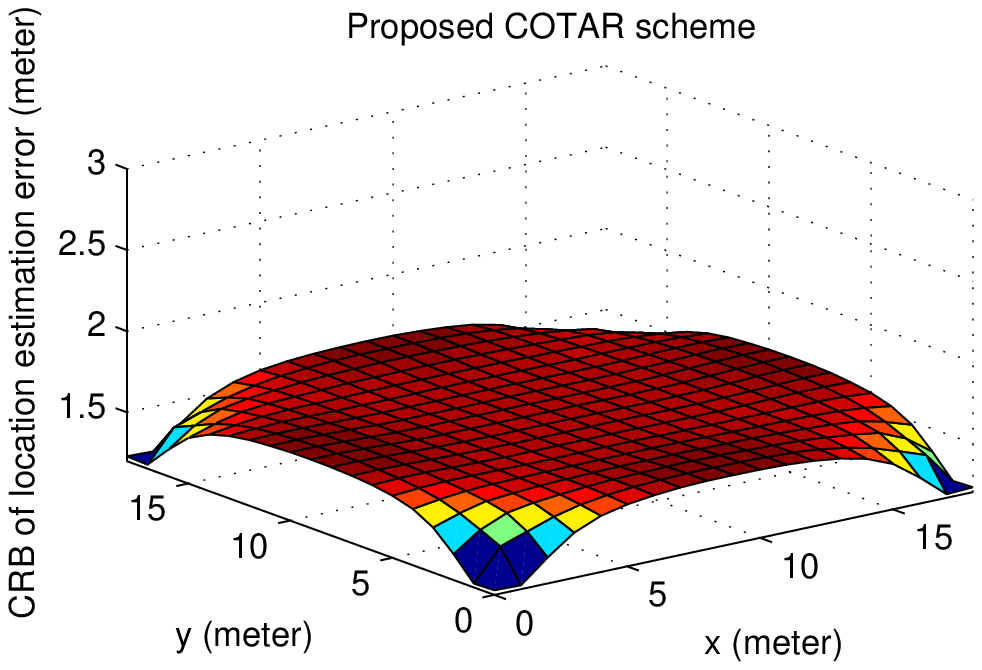}
\caption{The CRB for different localization schemes at different locations in clear LOS situations. Four reference nodes are deployed on the four corners of a 18 by 18 meters' square. For the proposed COTAR scheme, 4 target nodes cooperates localizing for each other in a 1 by 1 meter's grid.}
\label{fig:CRB_schemes}
\end{figure*}

The Cramer-Rao bounds for different schemes are shown in Fig.\ref{fig:CRB_schemes}, where 4 reference nodes are deployed at the four corners of an $18\times 18$ meters' square scenario in clear line-of-sight situation. Four cooperative target nodes help each other in the proposed COTAR scheme at the four corners of a $1\times 1$ meters' square. It is clear to see that the RSS only scheme has a poor localization accuracy from 8 to 10 meters. The traditional TOA scheme obtains a better accuracy of 2.7 meters in the middle and 2.9 meters at the four corners. The hybrid TOA-RSS scheme helps improve the accuracy of traditional TOA scheme at the four corners to 2.2 meters, but not in the middle of the scenario. The proposed scheme has an excellent localization accuracy, which achieves 1.55 meters' accuracy in the middle and 1.2 meters' accuracy at the four corners of the scenario.

\begin{figure}[htbp]
\centering
\includegraphics[width=5.in]{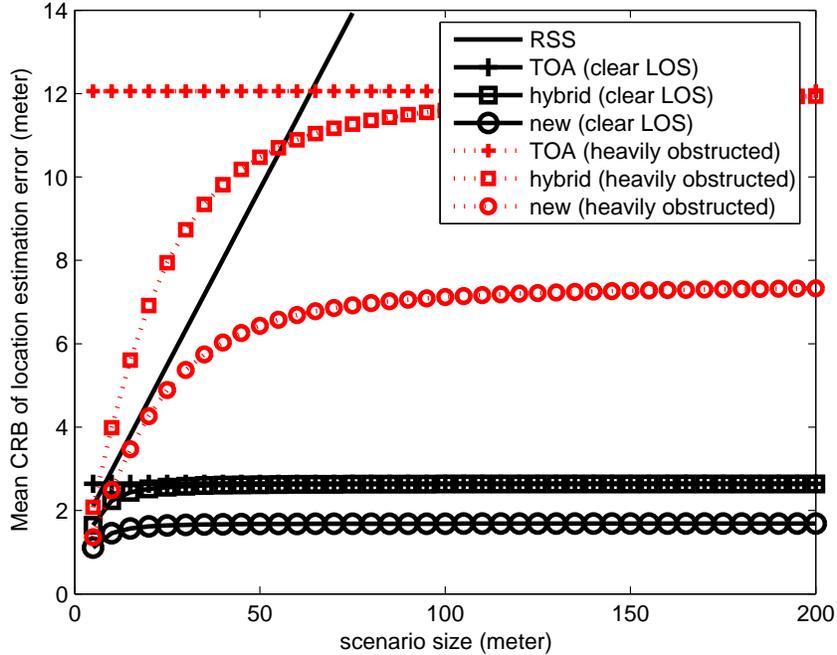}
\caption{The average CRB for different localization schemes vs increase of scenario size. Four reference nodes are deployed on the four corners of a increasing size's square. For the proposed COTAR scheme, four target nodes cooperate localizing for each other in a 1 by 1 meter's grid. Both clear line-of-sight and heavily obstructed situation are considered.}
\label{fig:CRB_vs_sizes}
\end{figure}

We further invest the localization accuracy vs increase of scenarios size for different localization schemes in different channel conditions in Fig.\ref{fig:CRB_vs_sizes}. It is shown that the RSS scheme's localization error increases linearly fast, which makes it useless in a large scenario. On the contrary, the accuracy of traditional TOA scheme does change with the increase of range. The hybrid TOA-RSS scheme can only improve the accuracy when the scenario size is less than 10 meters' in clear line-of-sight situation, and 100 meters in heavily obstructed situation. Otherwise, it can only obtains the same performance as traditional TOA scheme. For the proposed COTAR scheme, it performs excellently in any scenario size. It cuts half down the localization error from $12$ to $6.5$ meters in heavily obstructed scenario, and improves the accuracy from $2.7$ meters to $1.6$ meters in clear line-of-sight situation.

\subsection{Root Mean Square Error Analysis for ML Estimation}

We now analyze the performance error of the proposed detector in (\ref{eqn:estimator}), and evaluate the errors with respect to different positions $\langle\hat{\mathbf{x}}, \hat{\mathbf{y}}\rangle$.

The expression in (\ref{eqn:estimator}) shows that when the
measurement vector $\mathbf{r}$ is Gaussian distributed, the
estimation vector $\left(\hat{\mathbf{x}}\atop\hat{\mathbf{y}}\right)$ becomes a
$2N$ dimensional Gaussian random vector with a pdf:
\begin{align}
&p_{\mathbf{\hat{\bf x},\hat{\bf y}}}=
\frac{\!\exp\!\left\{\!\frac{-1}{2}\left(\mathbf{\xi}-\mathbf{E}\left[\mathbf{\hat{x}\atop
 \hat{y}}\right]\right)^T\mathbf{P}^{-1}\left(\mathbf{\xi}-\mathbf{E}\left[\mathbf{\hat{x} \atop \hat{y}}\right]\right)\right\}}{(2\pi)^{N}\!|\mathbf{P}|^{1/2}},
\label{eqn:pdf2}
\end{align}
where $\mathbf{P}$, the covariance matrix of the estimation vector
$\left(\hat{\mathbf{x}}\atop\hat{\mathbf{y}}\right)$, is defined as:
\begin{align}
\mathbf{P}&=\mathbf{E}\left[\left(\left( \hat{\mathbf{x}}\atop
\hat{\mathbf{y}}\right)-\mathbf{E}\left[\hat{\mathbf{x}}\atop
\hat{\mathbf{y}}\right]\right)\left(\left(\hat{\mathbf{x}}\atop
\hat{\mathbf{y}}\right)-\mathbf{E}\left[\hat{\mathbf{x}}\atop
\hat{\mathbf{y}}\right]\right)^T\right].
\end{align}
There are two types of inaccuracy associated with a Gaussian
distributed variable, the mean bias $\mathbf{\rho}$ and the variance
$\mathbf{\psi}$. The estimation error $\mathbf{e}$ should in general
consist of both:
\begin{align}
\mathbf{e}&=\left( {\hat{\bf x} \atop \hat{\bf y}} \right) -
\left({ \mathbf{x}\atop \mathbf{y}}\right)  = \mathbf{\rho + \psi}.
\end{align}

On the other hand, substituting (\ref{eqn:measurements}) into
(\ref{eqn:estimator}) and rearranging terms, we can write the
estimation vector as:
\begin{align}
\left(\hat{\bf x}\atop \hat{\bf y}\right)=&\left(\mathbf{x}\atop
\mathbf{y}\right)+(\mathbf{G}^T\mathbf{\Lambda}^{-1}\mathbf{G})^{-1}\mathbf{G}^T\mathbf{\Lambda}^{-1}\left[\mathbf{f(\mathbf{x},\mathbf{y})-f(a,b)-G\left[\mathbf{x}\!-\!\mathbf{x}_0
\atop \mathbf{y}\!-\!\mathbf{y}_0\right]+\mathbf{n}}\right] \label{eqn:rearrangment}
\end{align}

It is clear from (\ref{eqn:rearrangment}) that the estimation
error comprises both a linearizing error and a noise term. The
linearizing error in (\ref{eqn:rearrangment}) contributes to the
mean bias:
\begin{align}
\mathbf{\rho}\!=\!\!(\mathbf{G}^T\!\mathbf{\Lambda}^{-1}\!\mathbf{G})^{-1}\!\mathbf{G}^T\!\mathbf{\Lambda}^{-1}\!\!\left(\!\mathbf{f}(\mathbf{x},\mathbf{y})\!-\!\mathbf{f}(\mathbf{x}_0,\mathbf{y}_0)\!-\!\mathbf{G}\!\left[\!\mathbf{x}\!\!-\!\!\mathbf{x}_0
\atop\!\mathbf{y}\!\!-\!\!\mathbf{y}_0\!\right] \!\right).\label{eqn:linearization_part}
\end{align}
The estimator will be unbiased, if $\mathbf{f(x},\mathbf{y})$ is a linear vector function; but since it is not, the inaccuracy of the Taylor expansion in (\ref{eqn:Taylar_expansion}) thus introduces non-zero mean bias to the estimation. From (\ref{eqn:Taylar_expansion}), we also know that if the initial position $\langle\mathbf{x}_0, \mathbf{y}_0\rangle$ is sufficiently close to the actual position $\mathbf{x}_0\rightarrow\mathbf{x}$ and $\mathbf{y}_0\rightarrow\mathbf{y}$, the linearizing error $\rho$ in (\ref{eqn:linearization_part}) will become vanishingly small and can be safely ignored.

The residual noise in the estimation, caused by the measurement error vector $\mathbf{n}$, results in variation in the estimation results. This
noise part is expressed as:
\begin{align}
\mathbf{\psi}=(\mathbf{G}^T\mathbf{\Lambda}^{-1}\mathbf{G})^{-1}\mathbf{G}^T\mathbf{\Lambda}^{-1}\mathbf{n}|_{\mathbf{x}=\mathbf{x}_0,\mathbf{y}=\mathbf{y}_0}. \label{eqn:noise_part}
\end{align}
Here the error vector $\mathbf{n}$ subsumes all the error
contributors, including the measurement distortion of the signal TOA
and the RSS between the two cooperative nodes, and others
uncertainties in the system. It is reasonable to assume that the
measurement distortion $\mathbf{n}$ for TOA and RSS are unbiased,
such that $\mathbf{\psi}$ is a $2N$ dimensional column vector following a
zero-mean (vector) Gaussian distribution.

Since the mean bias is a constant vector given the initial position
$\left(\mathbf{x}_0\atop \mathbf{y}_0\!\right)$, the calculation for the
covariance matrix $\mathbf{P}$ of the estimation error is only
affected by (\ref{eqn:noise_part}):
\begin{align}
\mathbf{P}&=\mathbf{E}\left[\left(\psi-\mathbf{E}[\psi]\right)\left(\psi-\mathbf{E}[\psi]\right)^T\right]\nonumber \\
&=\left(\mathbf{G}^T\mathbf{\Lambda}^{-1}\mathbf{G}\right)^{-1}.
\end{align}
Note that the diagonal elements of $\mathbf{P}$ denote the variances
of the errors in the estimated positions.

Since the measurement distortion $\mathbf{n}$ is Gaussian
distributed with zero mean, the maximum likelihood or the least
square estimator for the linearized model is the same as the minimum
variance unbiased estimator. A scalar measure of the estimator
accuracy is the root mean square error $\epsilon$, which is defined
as
\begin{align}
\epsilon\approx\sqrt{\mathbf{tr}(\mathbf{P})/N}.
\label{eqn:asymptotic-RMS}
\end{align}

\begin{figure}[htbp]
\centering
\includegraphics[width=5in]{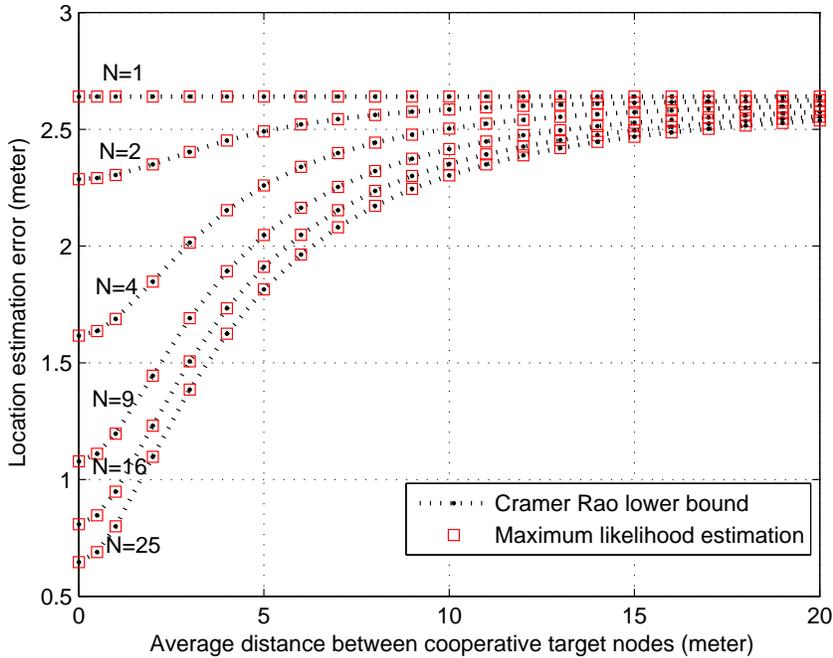}
\caption{Performance comparison between joint maximum likelihood estimation algorithm vs Cramer Rao lower bound for the proposed COTAR scheme under different number of cooperating nodes and distance.}
\label{fig:ML_vs_CRB}
\end{figure}

The RMS error bound of the proposed ML estimator for the COTAR scheme is plotted in Fig\ref{fig:ML_vs_CRB} as a function of the distance between target nodes and the number of cooperative target nodes. For comparison, we also shown the CRB of the COTAR scheme in the same picture. We consider four reference nodes are deployed at the four corners of a $1000\times 1000$ meters' square scenario, and target nodes are placed in a $(\sqrt{N}\Delta) \times (\sqrt{N}\Delta)$ meters' grid. The grid size $\Delta$ is the distance between two closest cooperative target nodes. It is well-known that the CRB sets the lower bound of a scheme for all the possible detectors, and may be achieved by an unbiased optimal detector. That our root mean square error bounds matches perfectly with the CRB (shown in Fig. \ref{fig:ML_vs_CRB}), and that our simplified iterative algorithm practically achieves these bounds (as will be shown in simulations), clearly indicates the efficiency of the proposed detection algorithm. We also find that the gird size $\Delta$ and the number cooperative target nodes are two important factors to the localization accuracy. For example, increasing the number of cooperating nodes from 1 (non-cooperative) to 25 can effectively improve the localization accuracy from 2.7 meters to 0.7 meters. It also shows that the cooperative target nodes should not be too far away. Due to the quickly-decreasing accuracy of RSS with distance, a close neighbor can do more to improve localization accuracy than several remote collaborators.

\section{Simulation results}\label{sec:simulation}
\subsection{COTAR vs Existing Strategies}
We now conduct computer simulations to study the realistic
performance for the proposed COTAR localization algorithm. We
firstly consider four reference nodes ($M=4$) are
located at the four corners of a square area with edge length $L=50$
meters.

To start, we consider four cooperating target nodes $(N=4)$
that move together in the square area and always stay
at the four corners of a $1\times 1$ meter's grid. Four localization
strategies  based on RSS, TOA, hybrid TOA-RSS, and the proposed COTAR, are evaluated:

\begin{figure*}[htbp]
\centering
\includegraphics[width=3.2in]{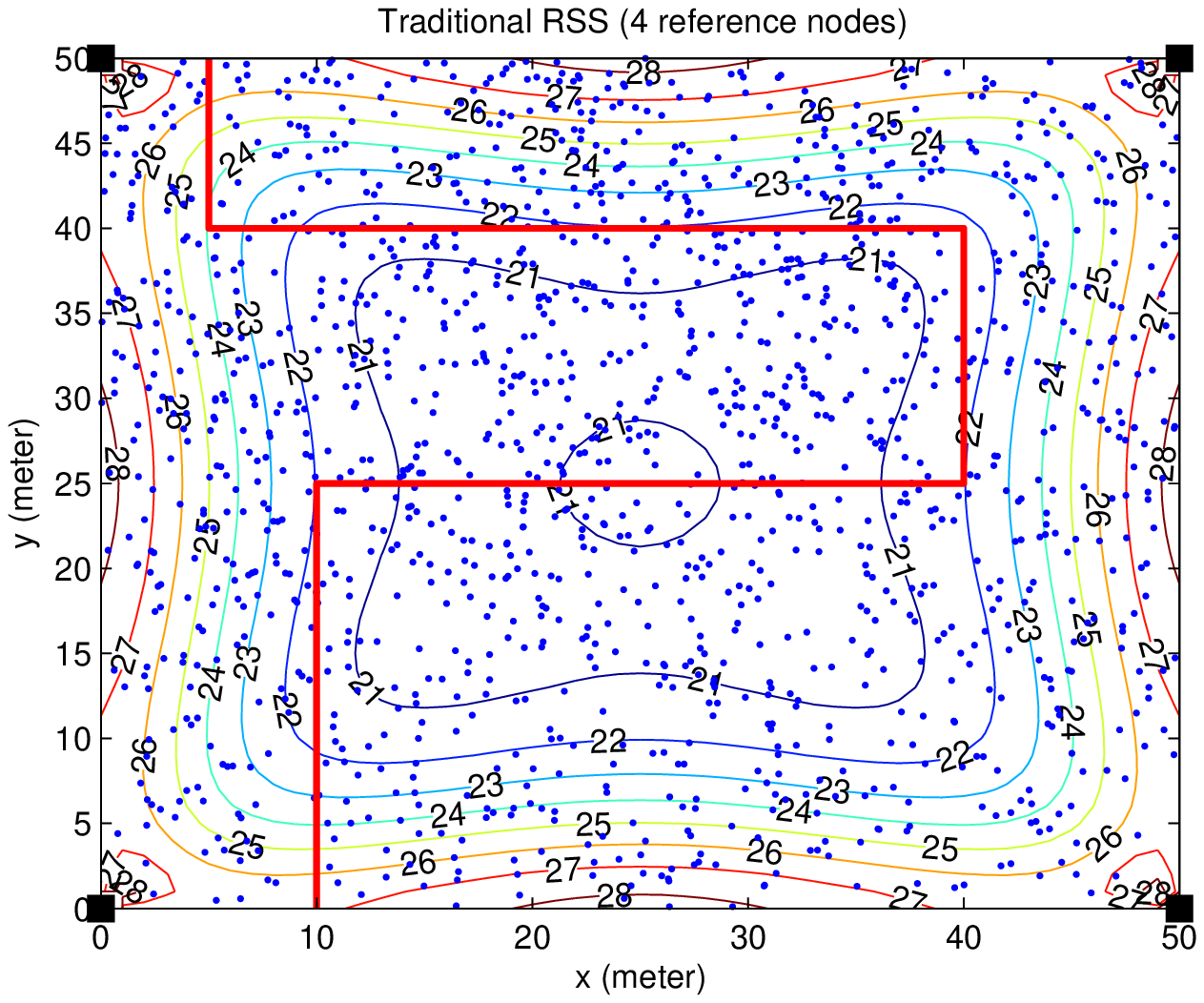}
\includegraphics[width=3.2in]{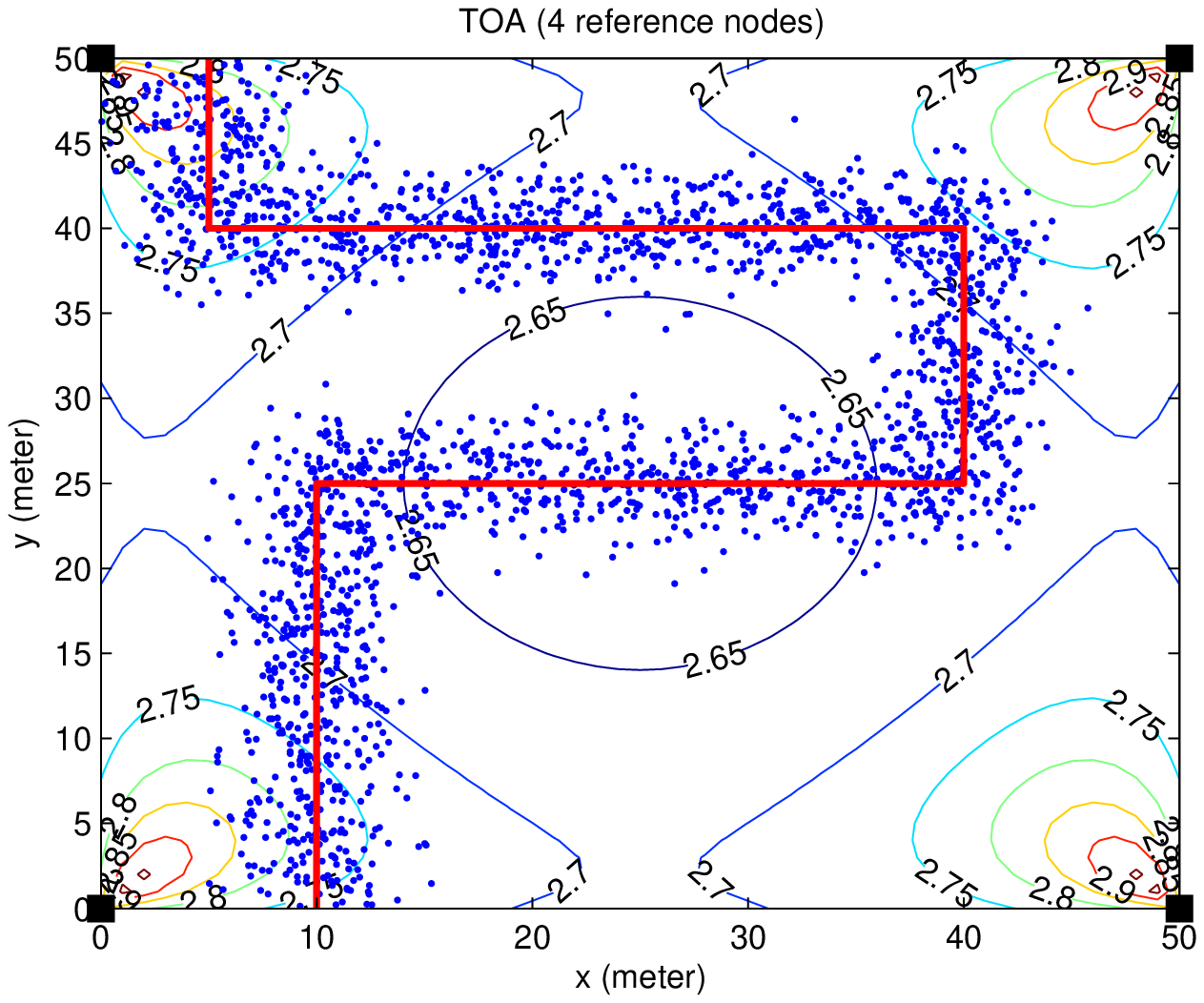}
\includegraphics[width=3.2in]{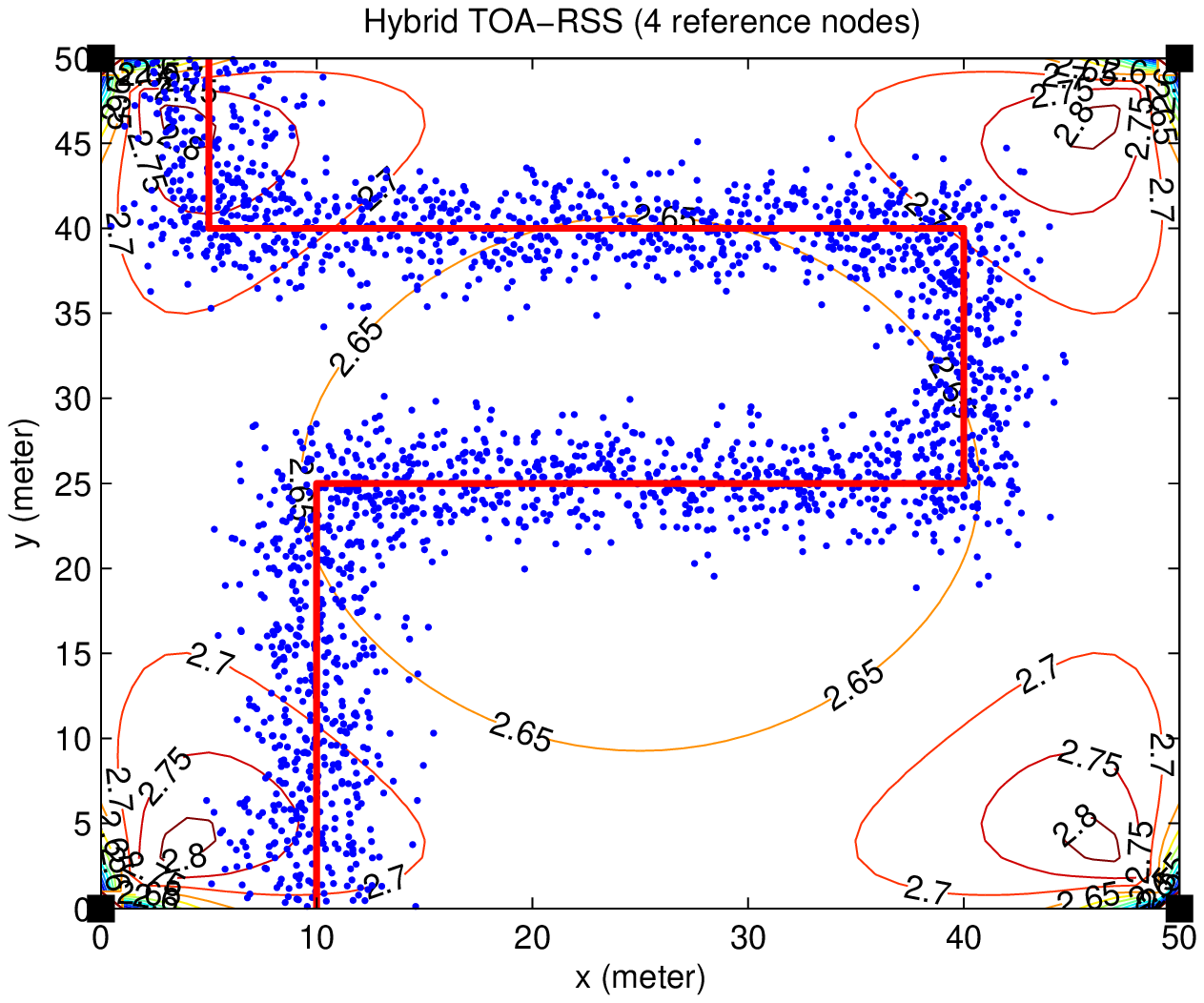}
\includegraphics[width=3.2in]{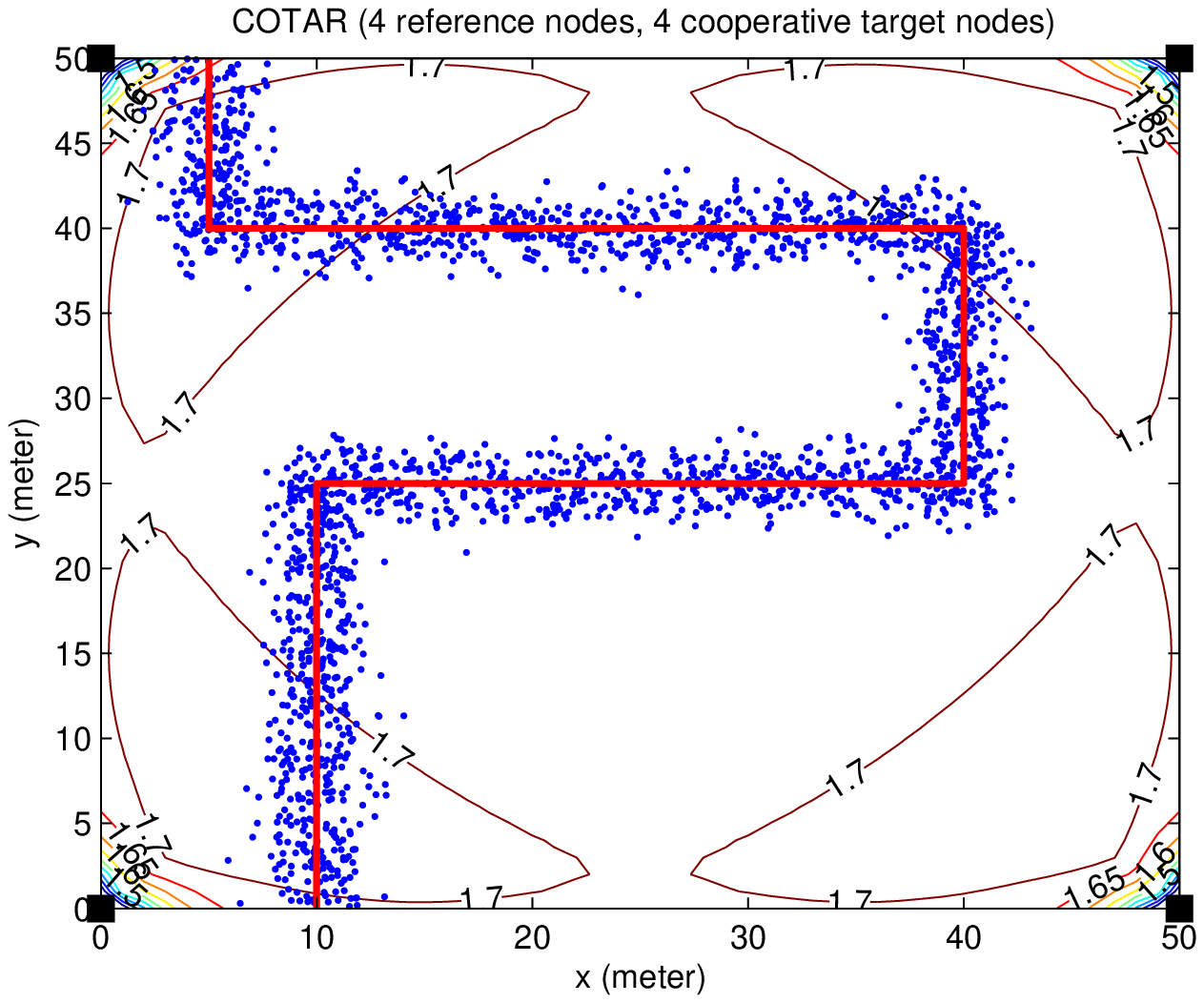}
\vspace{-0.5cm}
\caption{Impact of the number of cooperating nodes on the
performance of COTAR for different Rician distribution factor $K$.
$50\times 50$ meters grid, $M=4$ reference nodes at the corners,
bandwidth=2Mhz. }\label{fig:Simulation_comparison}
\end{figure*}

\begin{enumerate}
\item {\it RSS-only}: Each target node sends a beacon packet independently,
and all the $M$ reference nodes measure the RSS information, and
route them to a central node for maximum likelihood estimation. The
target nodes do not cooperate, and $M$ RSS measures are used to
estimate the position of any one target node.

\item {\it TOA-only}: It is similar to {\it RSS-only}, but the $M$ reference nodes
measure the TOA information instead. $M$ TOA  measures are
used to estimate the position of any one target node.

\item {\it Hybrid TOA/RSS }\footnote{For convenience, we call the RSS between target nodes and
reference nodes ``remote-RSS'' and the RSS between two (nearby)
target nodes ``neighboring-RSS.''}  \cite{TDOA-GRA}\cite{Hybrid_TDOA_AOA}: It is similar to {\it RSS-only}
and {\it TOA-only}, but the $M$ reference nodes measure both the
TOA {\it and} the RSS information. The scheme is again
non-cooperative, and  $2M$ measures ($M$ TOA and $M$ RSS) are used
to estimate the position of any one target node.

\item {\it COTAR}: As discussed before, the scheme is one that combines TOA and neighboring-RSS. A set of $N$ target nodes are engaged in a simple cooperation, and $MN$ TOA measures, $N \choose 2$ neighboring-RSS measure, and $NM$ remote RSS information are used to jointly estimate the positions of the $N$ target nodes. In the simulation, the scenario center with coordinates is always used as the initial position for both target nodes, and two iterations are used in the iterative refinement.
\end{enumerate}

The localization results for tracking mobile nodes are shown in Fig.\ref{fig:Simulation_comparison}, where red zigzag line shows the moving track of a group of four cooperative target nodes, and blue dots represents the estimated positions of different schemes. The results of 15 experiments are shown\footnote{We performed more experiments, but they make the picture too dense to see.}, and in each experiment, position tracking is performed in a continuous manner, with the previous estimated location serving as the starting point for the next one, and two iterations to estimate the incremental errors.
Also shown in the figure are the performance contour. Because of the large distance between each target nodes and the reference nodes, we are not surprised to see that the RSS-only method performs rather poorly, with the estimated positions scatter all over the place, yielding an estimation error RMS of as large as $10$ to $12$ meters. In comparison, the
traditional TOA-only method is much less sensitive to the distance. It attains a satisfactory performance of 2.7 meters of error. When TOA and remote-RSS are combined as hybrid TOA-RSS scheme \cite{TDOA-GRA}, only the performance
of the four corners of the scenario is effectively improved(, where reference nodes are deployed), but not the most of other positions especially in the middle of the scenario. The proposed COTAR strategy provides a much better solution by leveraging both the good accuracy of the RSS between two neighboring target nodes in short distance and the large localization range from TOA techniques. COTAR achieves a high localization accuracy of less than 1.5 meter's of distortion in
 the middle of the scenario, and 1.2 meters' accuracy at the four corners.

\subsection{Impact of Initial Position and Number of Iterations}

We next study the impact of the number of iterations and the initial
position on the localization accuracy. The same $50\times 50$ meters' square
area with $M=4$ reference nodes and $N=4$ cooperating target nodes
is simulated. We consider the clear line-of-sight channel situation,
the cooperative grid size to be 1 meter, and use the scenario center
as the initial position for the target nodes.

\begin{figure*}[htbp]
\centering
\includegraphics[width=4.5in]{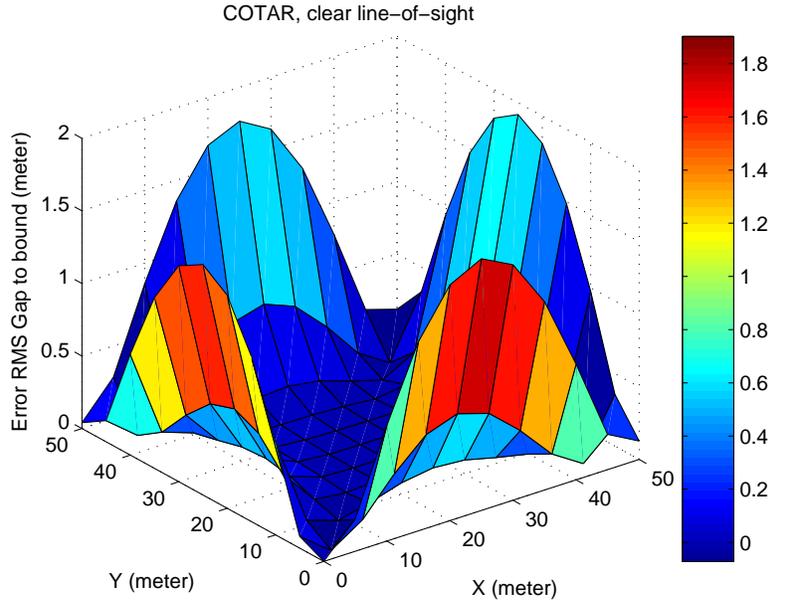}
\begin{center}(A)\end{center}
\centering
\includegraphics[width=4.5in]{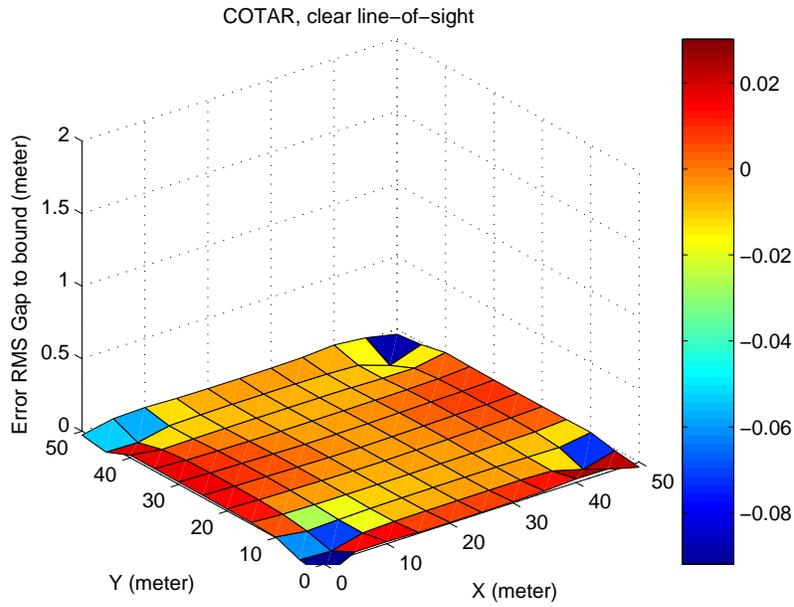}
\begin{center}(B)\end{center}
\caption{(A) RMS practice-theory gap (difference between the
simulated RMS errors and the theoretical lower bound) with only one
iteration. (B) RMS practice-theory gap with two iterations.
$50\times 50$ meters grid, $M=4$ reference nodes at the corners,
$N=2$ cooperating target nodes, bandwidth=2Mhz, Rician factor
$K=5$.}\label{fig:gap_to_bound}
\end{figure*}

We evaluate the gap between the actual localization error
and the theoretical lower bound. This gap is plotted in
Figure \ref{fig:gap_to_bound}(A) and Figure \ref{fig:gap_to_bound}(B) for
one and two estimation iterations, respectively. We see that the
accuracy of the starting point can make a difference, but only to
the first iteration. As shown in Figure \ref{fig:gap_to_bound}(A), with
only one iteration, the localization strategy may achieve a tiny RMS
error of $\le 0.01$ meter from the lower bound near the grid center,
but an accuracy gap of as large as 1.9 meters near the four edges.
It is interesting to observe that, when the target nodes get very
close to the corner points, their closeness to one of the reference
nodes, and hence the accurate TOA measures, can help mitigate the
negative impact caused by the inaccurate starting point (and the
insufficient iterations). As the iterations increases to two, the
accuracy gap quickly drops to near-zero  across the entire region.
In the simulated $50\times 50$ (meter) grid, it appears that when
the initial position is not offset by more than 10 meters from the
actual target locations, one iteration is enough; Otherwise, two
iterations suffice to bring down the estimation error to the lower
bound.

\subsection{Impact of Number of Reference Nodes and Cooperative Nodes}
\begin{figure*}[htbp]
\centering
\includegraphics[width=3.2in]{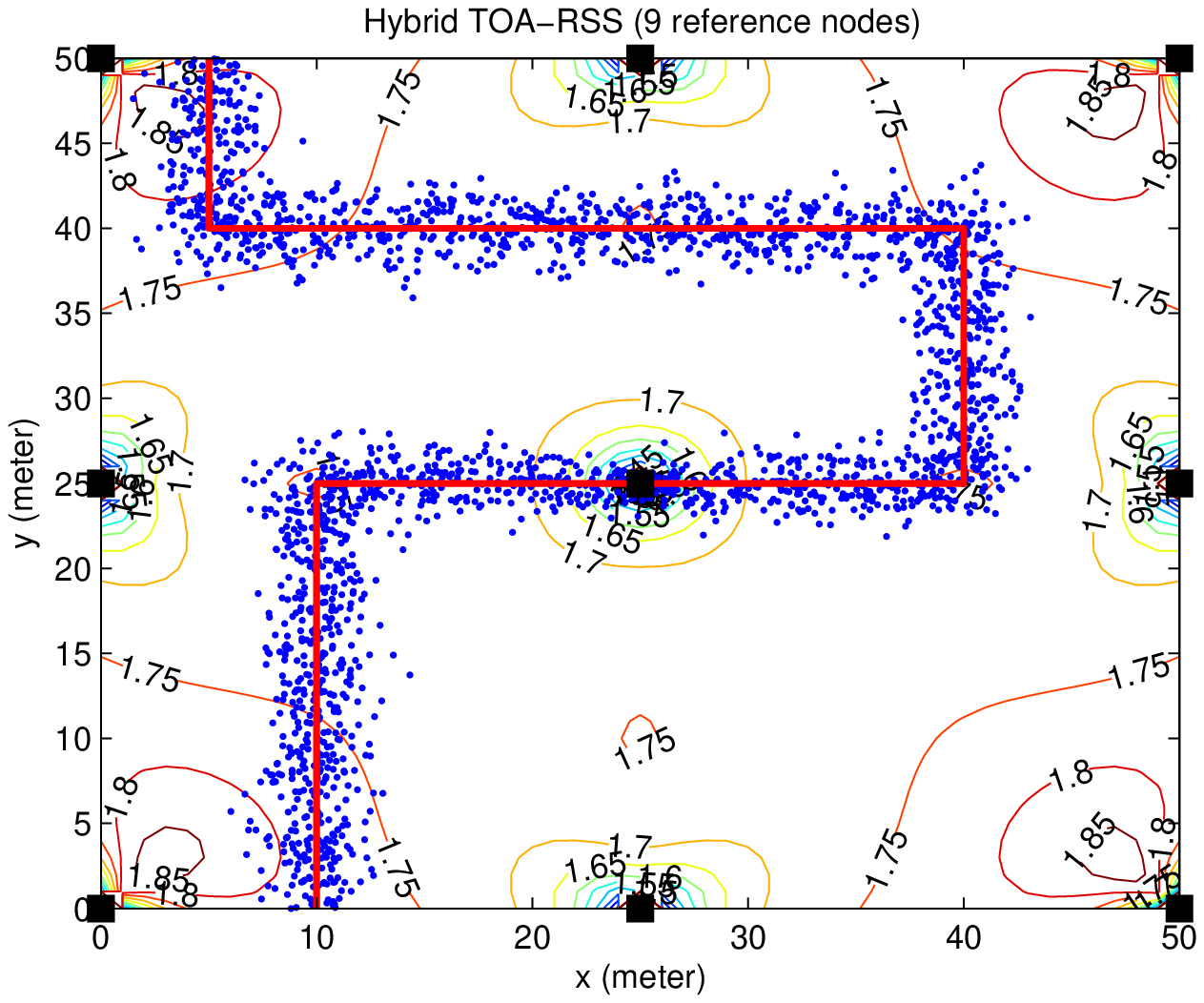}
\includegraphics[width=3.2in]{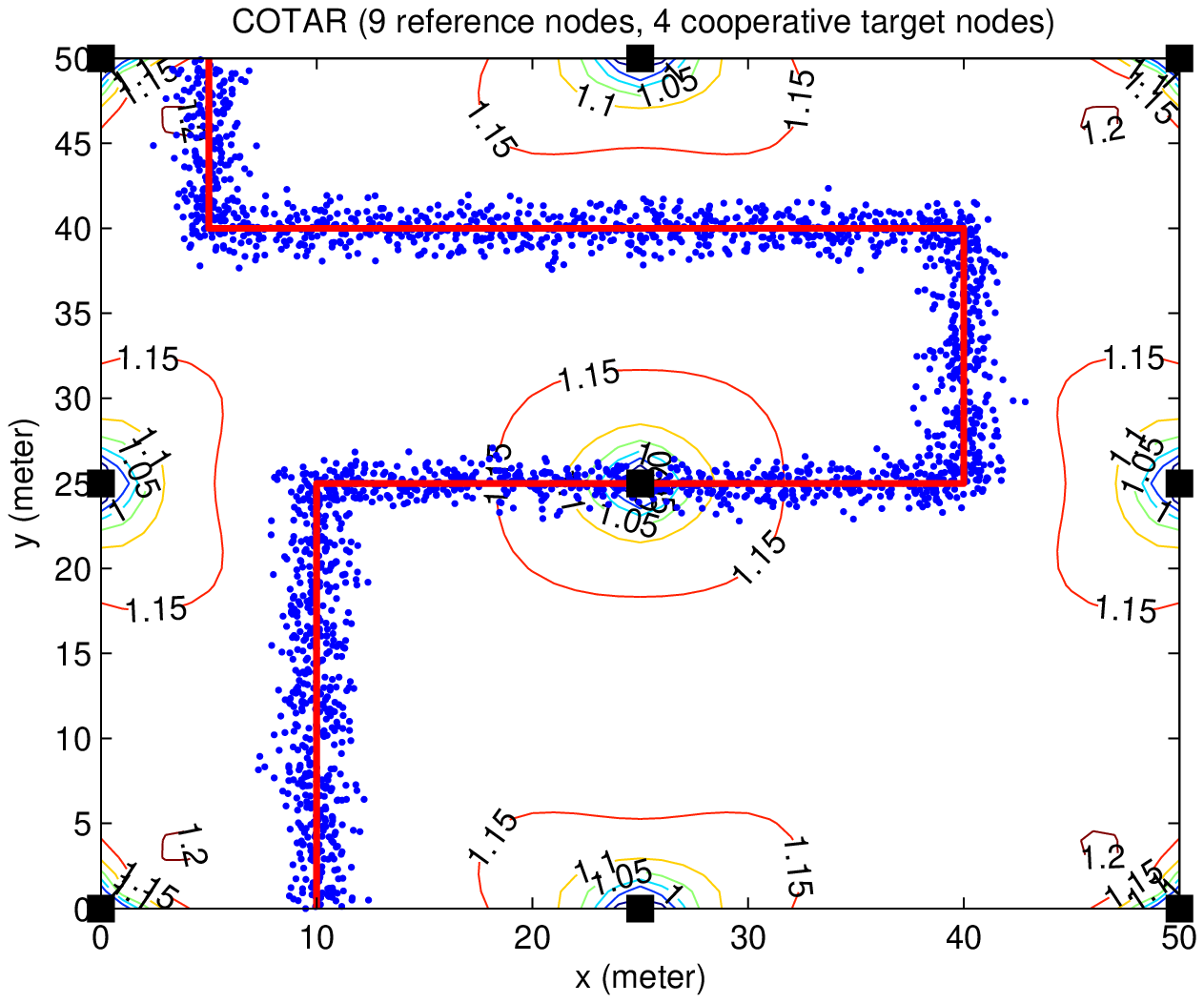}
\caption{Impact of increasing the number of reference nodes to 9 for the
performance of COTAR.
$50\times 50$ meters square scenario, $M=4$ reference nodes at the corners,
bandwidth=2Mhz. }\label{fig:simulation_9_references}
\end{figure*}

We first consider there are more than 4 reference nodes, say 9 reference nodes in the scenario. The 9 reference nodes are evenly deploy at every 25 meters as black $\blacksquare$ in Fig.\ref{fig:simulation_9_references}. We study the performance of the hybrid TOA-RSS scheme and the proposed COTAR scheme and plot the simulation results in Fig.\ref{fig:simulation_9_references}. From this figure, we see that the ML estimation algorithm works robustly in any position in the scenario, and increasing reference nodes number from 4 to 9 can effectively improve the localization accuracy for both hybrid TOA/RSS and COTAR schemes. At the middle of the scenario, the localization accuracy improve from 2.65 to 1.65 meters for hybrid TOA-RSS, and 1.7 to 1.05 meters for COTAR.

We further study the localization performance with more than 4
cooperative nodes. The same system set up is used, and we consider
the cases when $N=9 \mathrm{and} 16$ cooperative target nodes are
available for cooperation.

\begin{figure*}[htbp]
\centering
\includegraphics[width=3.2in]{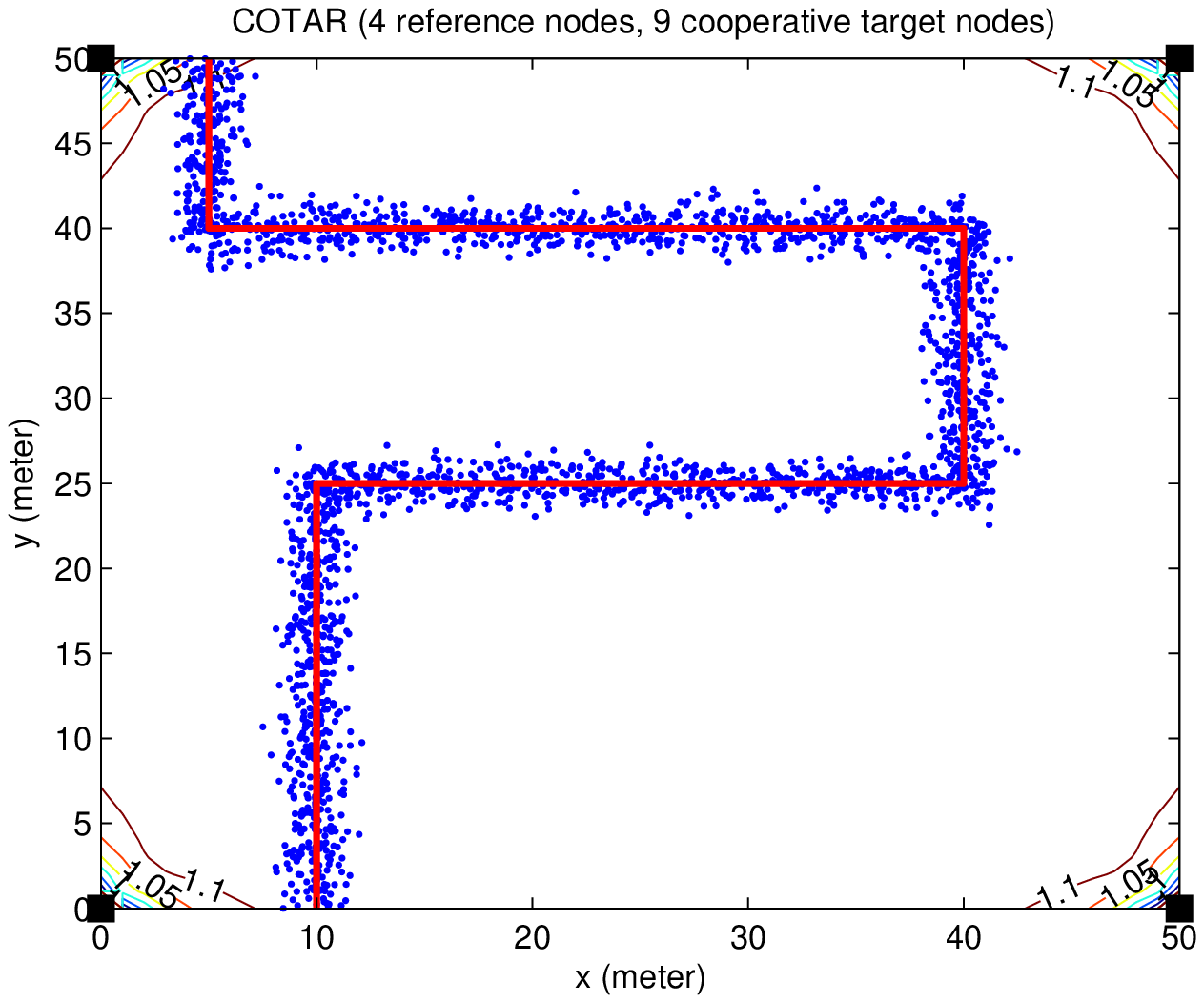}
\includegraphics[width=3.2in]{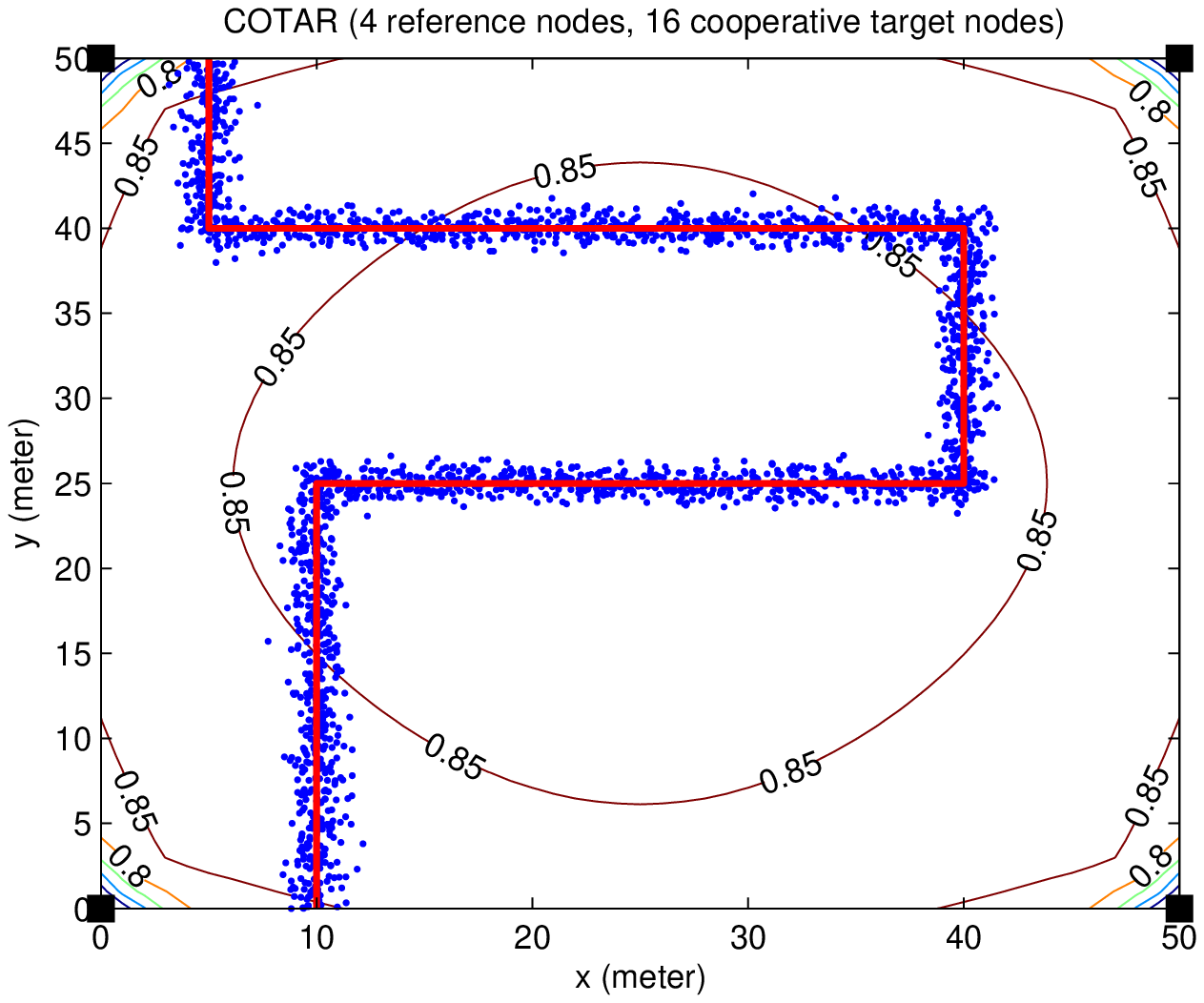}
\caption{Impact of increasing the number of cooperating nodes on the
performance of COTAR,
$50\times 50$ meters square scenario, $M=4$ reference nodes at the corners,
clear line-of-sight. }\label{fig:simulation_more_target_nodes}
\end{figure*}

The simulation results, plotted in Figure \ref{fig:simulation_more_target_nodes},
clearly speaks for the importance of increasing number of cooperative nodes.
With the help of 9 cooperating target nodes, COTAR can achieves 1.05 meters' accuracy
in most of the area. If we further increase the number of cooperative nodes to 16,
we are glad to find that COTAR obtains an excellent cooperative localization performance
of only $0.85$ meters' error.

\subsection{Impact of Missing RSS}
\label{sec:practical}

As mentioned before, although uncommon, it is possible for some RSS
information to miss in the equation, either because a target node
fails to estimate it in the first place, or because all the $M$
reference nodes fail to decode it from the beacon packet. In such
cases, the joint estimation should proceed without this information.
One possible way to handle such abnormality is to reformat all the
relevant matrices in (\ref{eqn:RSS_A}) by extracting the
corresponding row(s) from (\ref{eqn:RSS_A_pq}). A more convenient
way is to simply set the missing RSS information to be zero in the
original formulation in (\ref{eqn:RSS_A_pq}) without changing the
dimension of the partial derivative matrix.

\begin{figure}[htbp]
\centering
\includegraphics[width=5in]{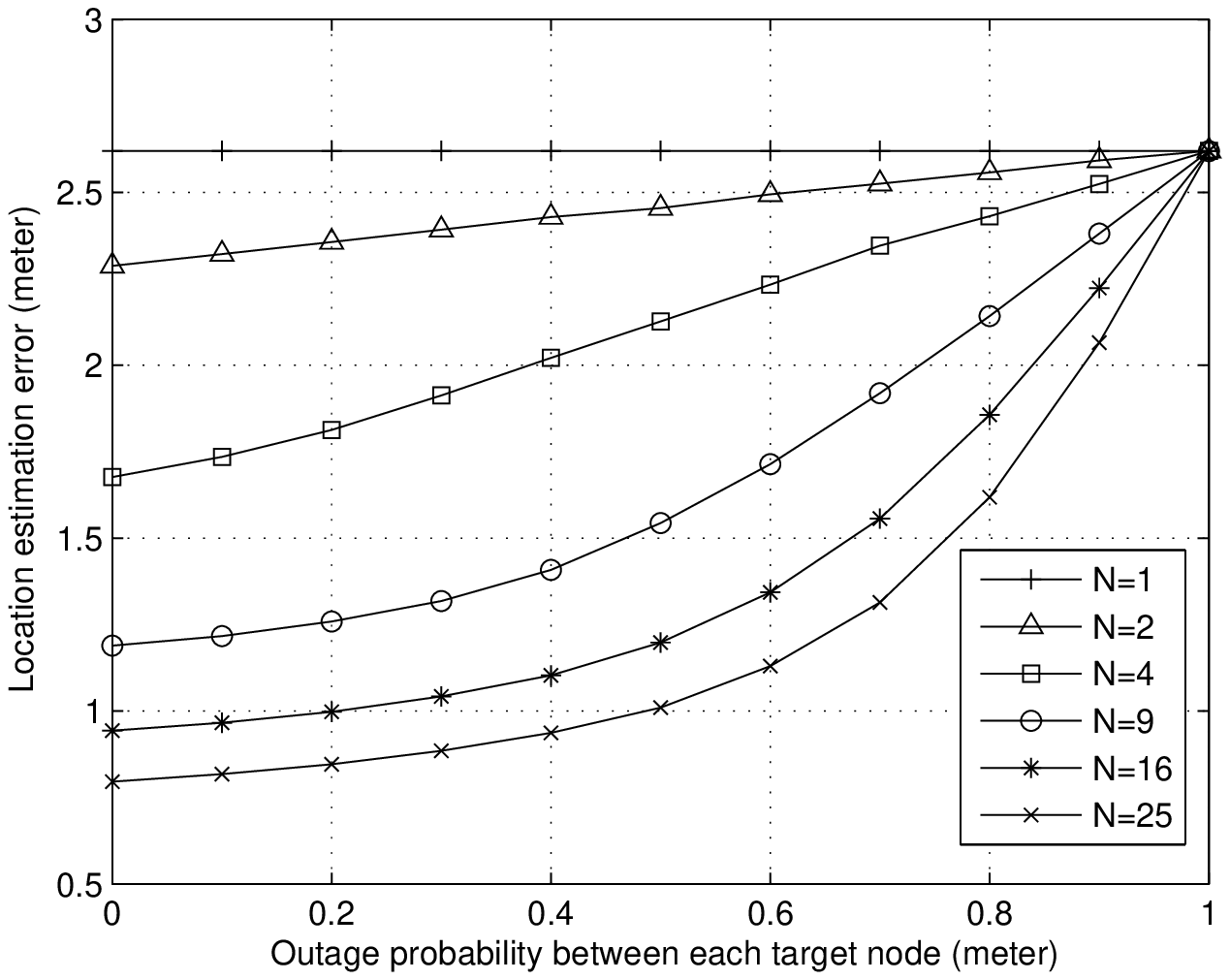}
\caption{Average localization accuracy as a function of the
percentage of missing (neighboring) RSS information. $50\times 50$
meters scenario, $M=4$ reference nodes at the four corners, $N=1, 2,4,9,16,25$
cooperating nodes, clear line-of-sight, target grid size to be 1 meter.}\label{fig:RMS_intra_outage}
\end{figure}

The potential impact of missing RSS information on the COTAR
performance is presented  in Figure \ref{fig:RMS_intra_outage}. Cases
with different cooperating nodes having different probability
(percentage) of missing RSS are evaluated. It is encouraging to see
that the localization accuracy degrades rather gracefully with
missing RSS. When the missing RSS is below 20 percentage, the
increase in average RMS error is very minor -- no more than 0.07
meters for $2$ cooperating nodes and negligible for $16$
cooperating nodes. When all the RSS is missing (100\% missing
probability), the COTAR strategy reduces to the non-cooperative
TOA-only scheme with RMS error of about 2.67 meters. It is also
evident from the simulation that increasing the number of
cooperating nodes is an effective way to compensate RSS loss and
reduce localization error.

\subsection{Tracking Highly Mobile Nodes}

The proposed localization strategy is also evaluated in a highly
mobile system such as the 3G mobile cellular phone networks. We
consider four base stations situated at the four corners of a square
grid with edge length $L=1000$ meters, acting as the reference nodes
($M=4$). Suppose a pair of cell phones are moving together in the
grid with a high speed of up to 160 km/hour. Suppose the cell phones move
with random directions and when they hit the boundary, they bounce
off like the reflection of a ray of light. The COTAR algorithm is
used to track the mobile positions of these cell phones, where the
previous estimate is used as the starting point for the next
estimate, and either one or two iterations of refinement is
performed. To make the system simple and light-load, we consider a
very low sample rate, such that beacon packets are transmitted and
localization is performed once every 5 seconds.

\begin{figure}[htbp]
\centering
\includegraphics[width=5in]{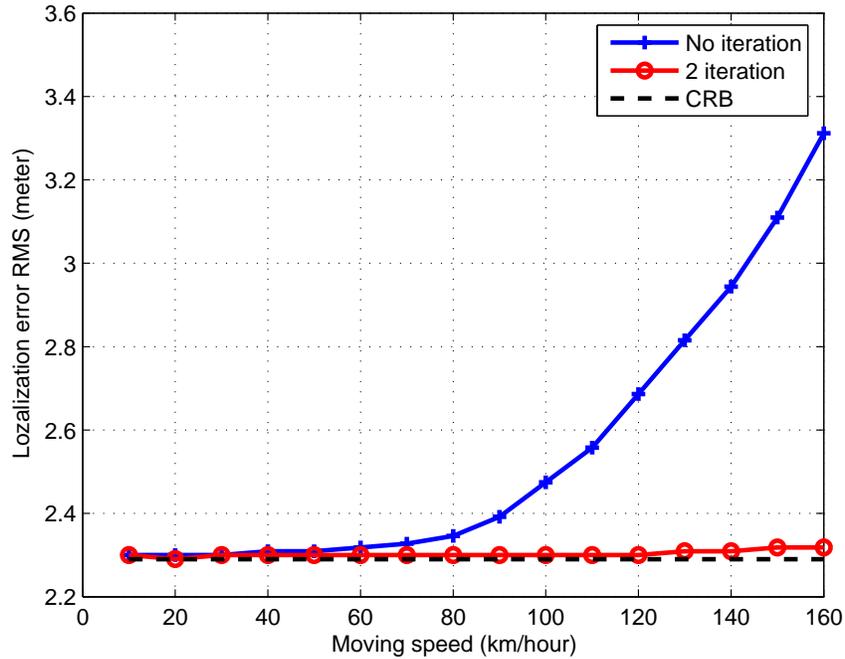}
\caption{Localization accuracy for mobile nodes at different moving
speeds and iteration numbers. $1000\times 1000$ meters grid, $M=4$
reference nodes at the corners, $N=2$ cooperating mobile nodes which are 1 meters' apart,
clear line-of-sight situation.}\label{fig:mobile_localization}
\end{figure}

The localization RMS error is plotted at a function of the mobile
speed in Figure \ref{fig:mobile_localization}. We can see that when the
nodes move at a slow to medium speed ($<80$ km/hour), 1 iteration is
enough to obtain good accuracy. As the mobile peed increases beyond
80 km/hour, since our sample rate is rather low and two succeeding sampling points
are fairly separated, and since the localization errors may accumulate,
1 iteration becomes a little insufficient, resulting an increase of
localization RMS error from 2.55 meters (at 80 km/hour) to 3.60
meters (at 160 km/hour). To bring down the error, one can simply add
one more iteration, such that the localization error is kept
consistently low and close to the lower bound.

\section{Conclusion}\label{sec:conclusion}

Low-cost and accurate wireless localization is challenging, due to
the limited bandwidth, hardware and computational resources, as well
as possibly severe multi-path fading and non-LOS conditions. In the
paper, we propose an effective cooperative-TOA-and-RSS localization
strategy which is simple, practical, and inexpensive to deploy, and
which requires no additional hardware than the usual cheap sensors
(like Zigbee nodes). By engaging two or more target nodes in simple
but effective cooperation, and by leveraging useful estimation and
detection techniques, the new COTAR strategy reaps the merits of
both TOA (i.e. decent performance fora a large localization range)
and RSS (i.e. high accuracy in short distances), without incurring
much cooperation overhead to the target nodes. Analytical and
simulation results confirm the good performance of the new strategy
in both clear and heavily obstructed cases. The proposed COTAR
strategy is particularly useful for (high-speed) mobile
localization, since the algorithm can naturally and efficiently
achieve incremental update through iterative refinement.

\end{document}